\documentstyle[12pt,aasms4]{article}
\received{09 June 1988}
\accepted{11 November 1988}
\input{psfig}

\slugcomment{ to appear in Ap.J.}
\lefthead{Flores et al.}
\righthead{15$\mu$m ISO$^1$ observations}

\begin{document}
\title{ 15$\mu$m ISO$^1$ observations  of 
  the  1415+52 CFRS field: the cosmic star formation rate
as derived from deep UV, optical, mid-IR and radio photometry}

\author{H. Flores$^{1,9}$, F. Hammer$^{1,9}$, T.X. Thuan$^{2}$,  
    C. C\'esarsky$^{3}$, F.X. Desert$^{4}$, A. Omont$^{5}$,  
      S.J. Lilly$^{6}$, S. Eales$^{7}$, D. Crampton$^{8}$ and 
O. Le F\`evre$^{1}$}

\affil{1) Observatoire de Paris, Section de Meudon, DAEC, 92195 Meudon 
  Principal Cedex, France.\\
  2) Astronomy Department, University of Virginia, USA.\\
  3) Service d'Astrophysique, CEA, France.\\
  4) Institut d'Astrophysique Spatiale, Orsay. France.\\
  5) Institut d'Astrophysique de Paris, France.\\
  6) Department of Astronomy, University of Toronto, Toronto, Canada.\\
  7) University of Cardiff, UK.\\
  8) Dominion Astrophysical Observatory, National Research Council of Canada,
  Victoria, Canada
}
\authoremail{flores@obspm.fr}

\altaffiltext{1} {Based on observations with ISO, an ESA project 
  with instruments funded by ESA Member States" (especially the PI 
  countries: France, Germany, the Netherlands and the United Kingdom) 
  with the participation of ISAS and NASA}
\altaffiltext{9} {Visiting Astronomer, Canada-France-Hawaii Telescope
(CFHT), which is operated by the National Research Council of Canada,
the Centre of Recherche Scientifiques of France, and the University of Hawaii.}

\begin{abstract}
The CFRS 1452+52 field has been deeply imaged 
with the Infrared Space Observatory (ISO) using ISOCAM through the LW3 filter   
(12-18$\mu$m). Careful data analysis and comparison to deep
optical and radio data have allowed us to generate a catalog of 78 15 $\mu$  
sources with both radio and optical identifications.
They are redder and lie at higher redshift than I-band selected galaxies,
with most of them being star-forming galaxies.\\

We have considered the galaxies detected at radio and 15$\mu$m wavelengths
which potentially include all strong and heavily extincted starbursts, up 
to z=1.
Spectral energy distributions (SED) for each of the sources have been derived 
using deep radio, mid-IR, near-IR, optical and UV photometry. The sources were 
then spectrally classified by comparing to SEDs of well known nearby galaxies. 
By deriving their FIR luminosities by interpolation, we can estimate 
their Star Formation Rate (SFR) in a way which does not depend sensitively
on the extinction.  75\% (-40\%, +10\%) of the star formation
 at z$\le$ 1 is related
to IR emission and the global extinction is in the range $A_{V}$=0.5 -- 0.85.
While heavily extincted starbursts, with SFR in excess of 
100 M$_{\odot}yr^{-1}$ constitute less than a percent of
all galaxies, they contribute about 18\% of 
the SFR density out to z=1. Their morphologies range from S0 to Sab, and more
than a third are interacting systems.\\

The SFR derived by FIR fluxes is likely to be $\sim$ 2.9 times higher
 than those previously estimated from UV fluxes.
The derived stellar mass formed since the redshift of 1 could be too high
 when compared to the present day stellar mass density. This might be due to 
an IMF in distant star-forming galaxies different from the solar 
neighborhood one, or to an underestimate of the local stellar mass density.
\end{abstract}

\section{Introduction}
\label{sec:intro}
 The 2800\AA\ and [OII]3727 emission line luminosity densities have decreased
by a factor of $\sim$ 10 from z = 1 to the present day (Lilly et al. 1996; 
Hammer et al. 1997). 
This has led Madau et al. (1996, 1998) to suggest that the cosmic star 
formation density has 
  decreased by the same factor within that redshift interval, and that 
most of the stars seen now were formed during the first half of
 the Universe's existence.
The UV
emission from galaxies is produced by a complex mix of short and moderately
long-lived stars, the latter (late B and A0 stars) contributing more at longer
UV wavelengths. Even old stars in luminous early-type and quiescent galaxies
 can contribute to the observed UV luminosity density. However 
the most important
uncertainty in estimating the star formation density from the 
UV luminosity density is due to the 
extinction which can show large variations from one galaxy to another. 
For example in IRAS star-forming galaxies, most of the energy is reemitted 
at far-IR (FIR) wavelengths and these objects are either missed or their star
formation rates are severely underestimated when derived by UV measurements.
 This is why it
is often thought that the UV luminosity density is likely to provide only
a lower limit to the actual star formation density. The
situation is complicated further by the expected contamination
by AGN to the UV light density.\\
In an attempt to better estimate the cosmic star formation density, 
Tresse and Maddox (1998) have
calculated the extinction-corrected $H\alpha$ luminosity density at z $\sim$ 
0.2.
Their result is in agreement with the UV (2800\AA) at z = 0.35 
(Lilly et al, 1996)
if an extinction of 1 mag is assumed for the UV continuum. A preliminary study
of more distant galaxies indicates that the situation might be similar at
z$\sim$ 1 (Glazebrook et al, 1998), but it is limited by the difficulty
of measuring the near-IR redshifted $H\alpha$ line of faint galaxies
with 4m telescopes. \\
Multi-wavelength analyses 
can provide, in principle, a detailed budget of the energy output in each  
wavelength range for the entire galaxy energy distribution. It has been shown 
 for local galaxies that FIR luminosities are tightly correlated
with radio luminosities (Helou et al. 1987; Condon 1992), and that bolometric
luminosities are most closely proportional to 12 $\mu$m luminosities
(Spinoglio and Malkan, 1989; Spinoglio et al, 1995). These trends hold
 over a wide range of galaxy luminosities, despite the large variety
of galaxy energy distributions. Only AGNs which are believed to be
associated to supermassive black holes, appear not to follow those relations
(Condon et al. 1988).\\
Recent observational advances allow now to study distant galaxies from the 
UV to the radio, sampling a wavelength range which covers most of the 
domain where their energy is being emitted.
 VLA deep surveys are able to detect sources down
to 10 $\mu$Jy (e.g. the 5GHz surveys of Fomalont et al. 1992
and Richards et al. 1998) and   
ISOCAM (C\'esarky {\it et al.} 1996) 
aboard the Infrared Space Observatory (ISO, Kessler {\it et al.} 1996)
 can reach detection limits
of 100$\mu$Jy at 15$\mu$m (Elbaz et al. 1998).
In the range 60--200$\mu$m the detection limits are 0.2 Jy at 60$\mu$m
 from the IRAS
Faint Source Catalog (Moshir et al. 1989) and 0.1 Jy at 175$\mu$m
from the FIRBACK survey carried out with ISO (Clements et al. 1998;
 Puget et al. 1998). FIR detections thus appear to be not sensitive enough
 to reach the same 
depth as radio and Mid-IR (MIR) deep surveys. 
For example, if we consider a strong
and highly reddened starburst (SBH in the terminology of Schmitt et al. 1998),
 a $S_{15\mu m}$=250 $\mu$Jy source would correspond to a 0.009 Jy source at 60$\mu$m, and a z=1 redshifted SBH with $S_{5GHz}$=16$\mu$Jy would
have 0.022 Jy at 175$\mu$m.\\
The sensibility and high spatial resolution of  
ISOCAM allow the study of distant 
field galaxies at MIR wavelengths (2 $\leq \lambda \leq $ 20$\mu$m).
Star-forming galaxies and AGN are easily detectable in the wavelength range
5-18$\mu$m, even at large distances (Franceschini 1991). 
The Canada-France Redshift Survey (CFRS)
 field at 1415+52 (Lilly et al. 1995a) is the second 
most observed field at all wavelengths after the Hubble Deep Field (HDF). 
While it does not go as deep, it is  
$\sim$ 18.5 larger in area and thus is more suited for source statistics when
a volume-limited (z $\le$ 1) sample is considered.
It has been  observed to very faint magnitudes in the BVIK bands (photometric 
completeness down to $I_{AB}$=23.5 mag, Lilly et al. 1995b), possesses 
spectroscopic data for galaxies brighter than $I_{AB}$ = 22.5 mag from 
the CFRS, and deep radio observations (S$_{5GHz} \geq$ 16 $\mu$Jy , Fomalont 
{\it et al.} 1992). 
The CFRS sample can be considered complete in the  sense that it  
 contains all luminous ($M_{B}(AB)\le$ -20.5) galaxies in the volume out to
 z = 1.\\
This paper presents a major follow-up study of the above CFRS field, 
by gathering and studying 
representative samples of galaxies selected at radio and MIR wavelengths.
With sensitivity limits of 250$\mu$Jy at 15$\mu$m (ISOCAM) and 16$\mu$Jy at 
5GHz (VLA), these
samples should include all strong and reddened starbursts up to z=1, 
with star formation
rates larger than 100 and 70 $M_{\odot}$ $yr^{-1}$ respectively. 
 These samples should not miss any luminous FIR source
in the CFRS field as the sources were  
selected using observations which cover wavelengths on
 either side of the 60-100$\mu$m bump.
They can thus be used ultimately to estimate the star formation density
 which has been missed by UV flux measurements.\\

 The nature of the $\mu$Jy radio sources in the field and of their 
optical counterparts
has been extensively discussed by Hammer et al. (1995). 
Recently this field has been imaged
by ISOCAM in the LW2 (5-8.5$\mu$m) and LW3 (12-18$\mu$m) filters.
We have presented the 6.75$\mu$m LW2 observations in a previous paper 
(Flores et al. 1998). There, we have discussed 
the details of the data reduction, the astrometry and the confidence 
level for each source. Fifty four sources with S/N $\geq$ 3 were detected 
with S$_{6.7\mu m}$ $\geq$ 150 $\mu$Jy, 21 of which possess spectra from the
 CFRS. 
Of the latter, 7 were stars. Among the 
non-stellar sources,  42\% were classified as AGN and 50\% as $S+A$ galaxies, 
i.e. star-forming galaxies with a significant population of A stars. The 
relatively high fraction of AGN is not unexpected because strong
 AGN are generally associated with a hot dust component and have generally
 bluer near-IR colors than starbursts. While the 6.75$\mu$m data appears
  not to be optimal for selecting starbursts, they are
useful for constraining galaxy SEDs, 
as galaxies (and especially Seyfert 2 galaxies) show large variations in their MIR 
color properties. 

We present here the 15$\mu$m LW3 observations. 
For galaxies in the redshift range 0.25 $\le$ z $\le$ 1,
 this filter samples the 8-12$\mu$ rest wavelength region. It can thus  
 simultaneously provide samples
of normal, starburst and active galaxies which are complete to a well-defined
 bolometric flux limit (Spinoglio et at, 1995). MIR measurements
are sensitive to dust thermal emission ($\lambda > 5\mu m$), its broad 
emission being interpreted as due to PAHs or Unidentified Infrared Bands 
(UIBs) carriers ($3\mu m < \lambda < 18\mu m$), and to nuclear 
non-thermal radiation. There have been recent suggestions that PAHs arise from
photo-dissociation regions around HII regions, the latter showing red
near-IR continua (Laurent and Mirabel, 1998, in preparation). Genzel
et al. (1998) have used the relative strengths of PAH features and near-IR
 colors
 to discriminate AGN dominated objects from starbursts. Lutz et al. (1998)
found that the fraction of AGN powered objects is relatively small at
moderate IR luminosities (typically $L_{IR} <$ 2 $10^{12}L_{\odot}$), but
reaches half at higher luminosities.
\\  
In section 2, we discuss the data reduction, the astrometry 
and the construction of the catalogs. In section 3 we present the 
 redshift distribution of the ISOCAM LW3 objects, and their optical
 properties including their morphologies from HST images. 
Section 4 presents spectral energy distributions from UV to 
radio wavelengths and a classification scheme for 
these and for radio-selected galaxies. Derivations of the UV and IR
luminosity densities are presented in Section 5. Section 6 describes the
global star formation density at z $\le$ 1 and compares it to previous 
estimates based on the UV luminosity density.

\section{Observations and  data reduction}
\label{sec:obs_and_red}
The CFRS field at 1415+52 was mapped with the ISOCAM LW channel 
(PFOV 6 arcsec
pixel) and the filter LW3 (12-18$\mu$m). Twelve individual images were 
obtained using the  micro-scanning AOT mode (CAM01) and resulting in a total 
integration time of $\sim$ 1200 sec pixel$^{-1}$. 
The micro-scanning mode provides the best
spatial resolution by superposition of images.  The same pixel of
the sky was placed in different parts of the camera in order to
minimize and detect any systematic effects. The micro-scanning AOT
technique also allows an accurate flat-field image to be generated and
yields a pixel size of 1\farcs5 in the final integrated image.  The
detection and removal of transients and glitches, integration of
images, absolute flux calibration, and source detection were carried
out using the method described by D\'esert et al. (1998).  This method
has been found to be
particularly well adapted to our observational strategy, i.e. coadding
the twelve images, without redundancy within each image.  Special
attention was paid to possible error propagation in the flux values.
 The photometric 
accuracy has been discussed by Desert {\it et al.} (1998). From the
 stellar energy
distributions, we find the photometry to be accurate to a few percent 
for sources with S/N $>$ 10, and to $\sim$50\% for sources with S/N = 3.\\
 Figure 1 displays the final 15$\mu$m image of the CFRS 1415+52 field. 
Individual images were carefully registered with each other in order to
optimize the image quality of the brightest compact objects (see Flores et al,
1998).
 The final image of the whole ISO field has a resolution equivalent
to a median FWHM$\sim$11\arcsec\ (calculated with DAOPHOT under IRAF).
It can be seen that the noise structure (Figure 1) is relatively homogeneous 
(the standard deviation is lower than one tenth of the mean) in $\sim$ 85\% of 
the area of the image, except near the edges.\\

\clearpage

\begin{figure}
\begin{center}
    \leavevmode
    \psfig{file=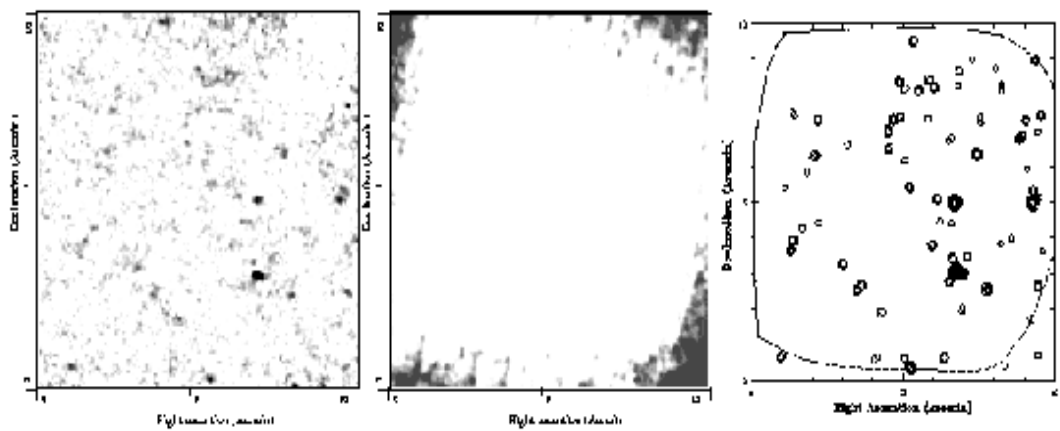,height=9cm,width=15cm}
\caption{ At left is the combined LW3 (12-18$\mu$m) image of the 
  whole 10'x10' CFRS 1415+52 field. Center coordinates are 
$\alpha$(2000)= 14 17 53.7 and $\delta$(2000)= 52 30 30.7. 
The scale is 1.5'' per
  pixel. The image has a FWHM resolution of $\sim$11\arcsec 
  ~and the scale is 1\farcs5 per pixel. In the middle is shown the
map of noise. In 85\% of the map, the noise is within 10\% of the average.
At right are shown the locations 
  of the catalogued 15$\mu$m sources which have successfully passed 
  our selection criteria (see text). Sources with S/N= 3 
  are represented by a single circle, while sources with S/N= 4, 6 and 8 
  are shown by 2, 3 and 4 concentric circles. }
  \end{center}
\end{figure}
\clearpage

 Point sources are iteratively extracted with a Gaussian PSF of 9 
arcsecond FWHM; a correction factor of 1.39 (deduced from a detailed 
modelling of the effective ISO PSF, Cesarsky {\it et al.}, 1996) is 
applied to the measured flux (calibrated with ISOCAM user's manual 
conversion table) to account for losses in the wings of the true PSF.

\subsection{ISO source catalogues}
\label{sec:iso_cat}
Source detections were made on the basis of S/N and repeatability in three 
independent combinations of the 12 individual images (for details about 
the source detection repeatability and classification, see F.X.
 Desert  {\it et al.}  1998). 
The repeatability test is based on the redundancy factor,
which is the number of times that the sky pixel was seen by different
pixels on the camera. The software built three independent projection
subrasters, and for each source candidate the flux and error are measured 
at the same position in each subrasters.
The quality factor is based on flux measurements and varies from the 
best confidence index (= 4) to the worse confidence level 
(= 0, see eqs. 7 to 11 of D\'esert et al.,1998). We have considered 
only sources with a 
confidence level higher than 3, which means that the final source flux is
 within 3 $\sigma$  (2 $\sigma$ in the 
case  of 4) of the source fluxes in subrasters,  
where $\sigma$ is the error in the final source flux.\\

 Altogether, 78 sources with S/N $\ge$ 3 fulfill these detection criteria 
(Table 1).
 We have considered as secure detections those sources with S/N $\ge$ 4 
(41 sources, catalogs 1, 2 and 3). The 37 sources listed in catalogs 4, 5 
and 6 with 3 $<$ S/N $<$ 4 are considered ``less'' secure.
 That a S/N $\geq$ 4 is a good detection criterion 
 is confirmed by studies in the Lockman Hole Deep Survey 
(C\'esarsky et al. 1998, in prep) which show that
  sources with S/N $\geq 4$ 
in individual frames are confirmed in 95\% of cases in the final integrated
 image (S/N $>$ 10). 

\begin{figure}
\begin{center}
    \leavevmode
    \psfig{file=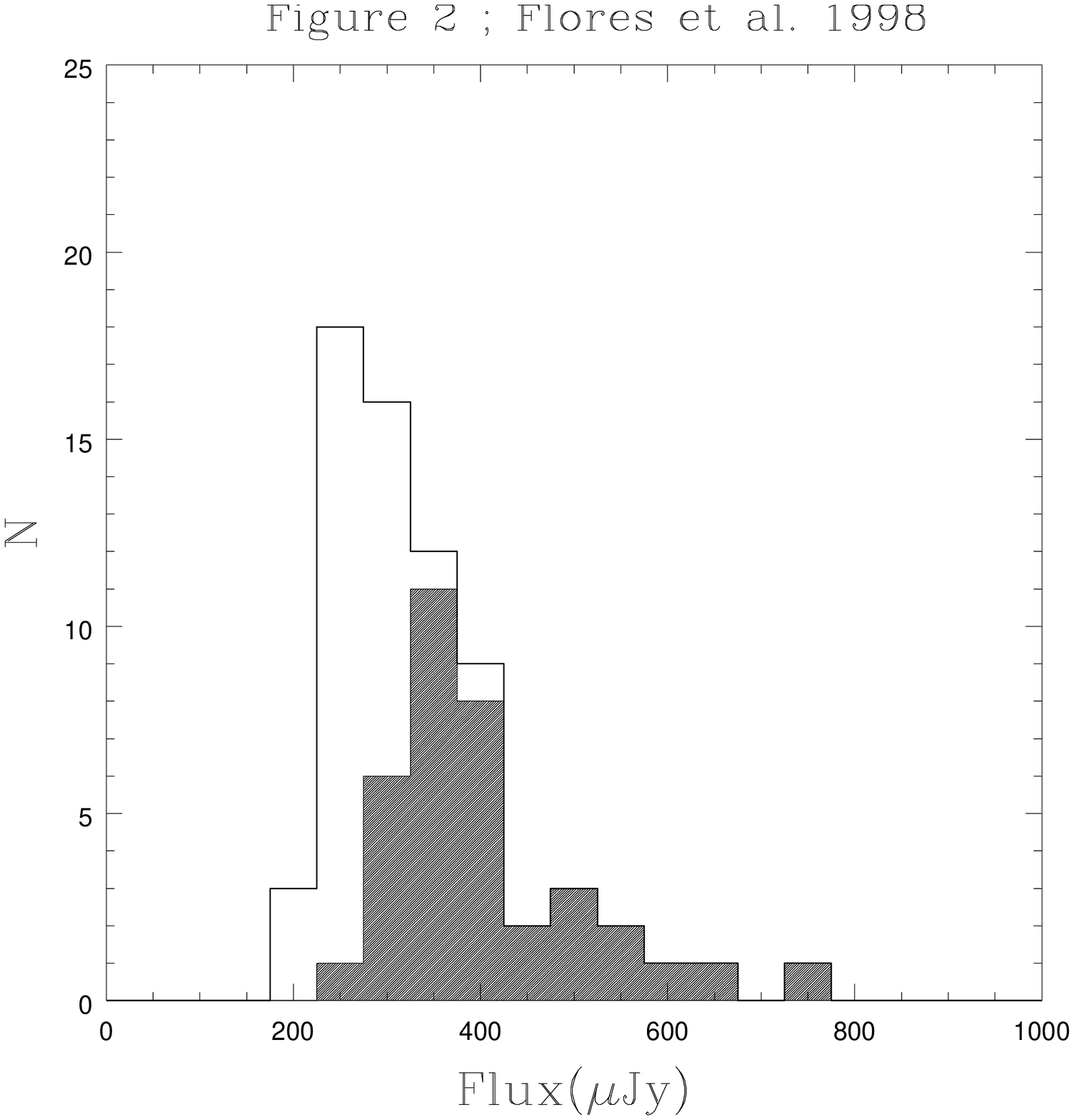,height=12cm,width=12cm}
\caption{ The flux distribution of ISOCAM 12--18 micron sources. 
  Sources with S/N$>$ 4 (catalogs 1 \& 2) are shown by the shaded histogram.}
\figcaption[fig3.eps]{(I-K)$_{AB}$ color distribution of 15$\mu$m sources 
  with $I_{AB} \le$22.5 (bottom panel) 
  compared to the CFRS galaxy color distribution (top panel). }
  \end{center}
\end{figure}

Figure 2 shows the flux
 distribution and the comparison to results from other fields suggests a 
completeness down to $ \sim$ 350 $\mu$Jy for 
sources with S/N $\geq$ 4, and to $\sim$ 250 $\mu$Jy for sources with 
S/N $\geq$ 3, at least for the central 85\% of the 10'x10' field.
These correspond to number densities of 1590 sources per square degree,
for $S_{15\mu m} \ge$ 350 $\mu$Jy, slightly larger than 
the number density of 1260 source per square degree found by 
Elbaz et al. (1998) in their ISOCAM survey of the Hubble Deep
Field.\\
Surveys of low S/N ratio sources can be affected by several biases,
the main ones being the possible unreliability of sources near the survey 
limit and the Eddington bias. 
Near the flux density limit, completeness falls off and  
possible false sources may be introduced. The fraction of false sources 
can be controlled through a random match
control test based on the identification rate of ISOCAM sources 
with optical sources (see next section) and should be small. 
We have checked whether our higher number count relative to the HDF data 
could be due to the Eddington bias.
 Our survey is intermediate in depth between the
Lockman Hole Deep survey (from 450 to 1400 microJy) and
the Hubble Deep Field survey (from 125 to 350 microJy).
 Elbaz et al. (1998) quote a slope of 1.3 and 2.1 for the HDF and Lockman Hole
respectively, and we should expect an intermediate slope near $\sim$ 1.7, lower
than the value of 2.2 derived from our S$\ge$250 and S$\ge$350 $\mu$Jy 
counts. 
We have attempted to estimate the fraction of sources which may have entered
 our sample erroneously because statistical uncertainties 
bring them over the detection threshold, 
while they actually possess a lower flux density than S=250 microJy. We 
have adopted a slope of 1.7 and perform 
 Monte Carlo simulations of 30 000 sets, 
 assuming a gaussian distribution for the
noise of faint sources. The simulations show that $\sim$ 15\% of the 
sources with S/N $\ge$ 3
can be sources with fluxes lower than 250 microJy, but that fraction drops to
only $\sim$ 2\% for sources with S/N $\ge$ 4.
This is probably close to the truth since after accounting for the Eddington 
bias (i.e. removing 15\% of the 78 sources with S/N $>$ 3 as well as 2\% of
the 41 S/N$>$ 4 sources), the slope becomes 1.7,  in good
agreement with other surveys.\\

Biases in source counts thus affect mainly the faintest sources in
our sample (those with 3 $\le$ S/N $<$ 4). They are the least powerful sources,
and as such cannot introduce significant uncertainties on 
global luminosity density estimates.

\subsection{ Astrometry and counterparts at radio and optical wavelengths}
\label{sec:astrometry}
The positional accuracy of ISO sources is affected by both the pixel size and  
the distortion (see Flores et al, 1998). The superposition of the ISOCAM field 
onto the CFRS 1415+52 optical field was done by matching the positions of the 6 
brightest sources (which were also detected at 6.75$\mu$m). 
By comparing the astrometry of the optical and ISOCAM LW3 sources,
we find a median difference of $\sim$3.7 arcsec. This is reasonable given that
the ISO pixel size is $\sim$6.0 arcsec. \\

The astrometric accuracy of the CFRS 1415+52 optical field is 0''.15, 
based on the comparison between the optical and
radio frames (Hammer  {\it et al.}, 1995). We first compare the 15$\mu$m image 
with the VLA $\mu$Jy radio map (Fomalont et al, 1991), and calculate the 
probability of a pure coincidence, assuming Poisson statistics: \\
\begin{equation}
P(d, S_{5GHz}) = 1 - e^{-n(S_{5GHz}) \pi d^2} 
\end{equation}

where d is the angular distance between the ISOCAM LW3 source and the radio
source in degrees and n is the integrated
density of radiosources with flux $S_{5GHz}$ ( $dn(S_{5GHz}$) = 83520 
$S_{5GHz}^{-1.18} dS$). Ten
ISOCAM LW3 sources are thus identified (see Table 1). They all possess optical
counterparts (Hammer {\it et al.} 1995). \\

Astrometry of ISOCAM LW3 sources not detected in radio has been derived 
by comparison to the optical sources in the CFRS 1415+52 field, and

\begin{equation}
 P(d, I_{AB}) = 1 - e^{-n(I_{AB}) \pi d^2} 
\end{equation}

is the probability of a pure coincidence, where $n(I_{AB})$ are integrated 
counts derived from the CFRS. In table 1, six catalogs
 have been defined with different S/N values 
($3 \leq S/N \leq 4, S/N > 4$) and P values ($P<0.02, P>0.02$, 
and sources without optical counterparts within 12 arcsec).  Objects which 
are also radio sources are indicated. These all have very low 
probabilities of accidental coincidence.\\

We also compare the 15$\mu$m map with that described in Flores et al. (1998) at
6.75$\mu$m. We found 17 of the 78 15$\mu$m sources to be also
detected at 6.75$\mu$Jy (with flux densities above 150$\mu$Jy).
 The fraction of 15$\mu$m sources also detected at 6.75$\mu$m increases  
with S/N (see Table 2).\\

The final noise structure in ISOCAM images 
is not strictly gaussian, because of possible 
residuals glitches, and because the 15$\mu$m positions could not have
 been fully corrected for image distortions. To calibrate the 
probabilities given in Table 1 in an empirical
 way,  we have applied a  random match control test to the ISOCAM image  
by rotating it successively by
45, 90, 180 and 270 degrees relative to the optical image. Only 3$\pm$1
of the 78 ISOCAM sources were found to be randomly associated with an optical
counterpart with $I_{AB}<$22.5 and P $<$ 0.02. This should be compared to 
the 37 $I_{AB}<$22.5 and P$<$0.02 counterparts found in Table 1. These 
experiments suggest
 that non-gaussian effects cannot affect our probability calculations by 
more than a factor 2.\\

Table 1 lists the sources in six catalogs, 
the confidence level decreasing in each successive catalog.   
The ISOCAM LW3 sources and their optical counterparts 
are given in columns (1) \& (2). Column (3)
provides the optical source redshift when available, " star " 
indicates that the ISO source is stellar, while " --- " indicates 
that no redshift is available.  In columns (4), (5) and (6) are the I, 
V and K isophotal magnitudes in the AB system, and " -- " indicates
that no photometry is available. Column (7) gives the angular 
distance in arcsec between the ISOCAM LW3 source and optical (or radio) 
identification,  while column (8) gives the associated 
probability of coincidence. Columns (9) and (10) 
give the flux density at 12--18  $\mu$m and its error in $\mu$Jy.
The positions and flux densities of the 9 non-identified ISO sources are
also listed in Table 1 (catalogues 3 and 6).

Among the sources with $S/N \ge$ 4, only 8 
(19\%) have counterparts fainter than  $I_{AB}$ = 22.5 , or no optical 
counterparts. These sources could be at redshifts higher than 1. 
The fraction of very faint optical counterparts  increases for the fainter
 ISOCAM LW3 sources (41\% for sources with 3 $<$ S/N $<$ 4), which is to 
be expected. On the other hand, in the most uncertain catalog (catalog 5), 
one can expect  $\sim$ 4.6 sources out of 19 to be pure coincidental 
projection. In seven  cases, more than one  optical counterpart appear to be 
related to  an ISOCAM LW3 source. In the following analyses, we have used the
 " best " optical identification (i.e. ,  those with lowest coincidental 
probability), but in Table 1, we have listed all possible optical
 identifications with a probability within a factor two of the smallest 
probability. \\

\section{Optical properties of the 15$\mu$m counterparts}
\label{sec:analyses}
\subsection{ Color and redshift distributions}
\label{sec:count_and_color}

The $(I-K)_{AB}$ color distribution of the 15$\mu$m optical counterparts 
(Figure 3) is significantly redder than that of the CFRS survey 
(top panel), with a median 0.5 magnitude redder than that of
 CFRS galaxies ($<(I-K)_{AB}>\sim 1.3$). 

\begin{figure}
\begin{center}
    \leavevmode
    \psfig{file=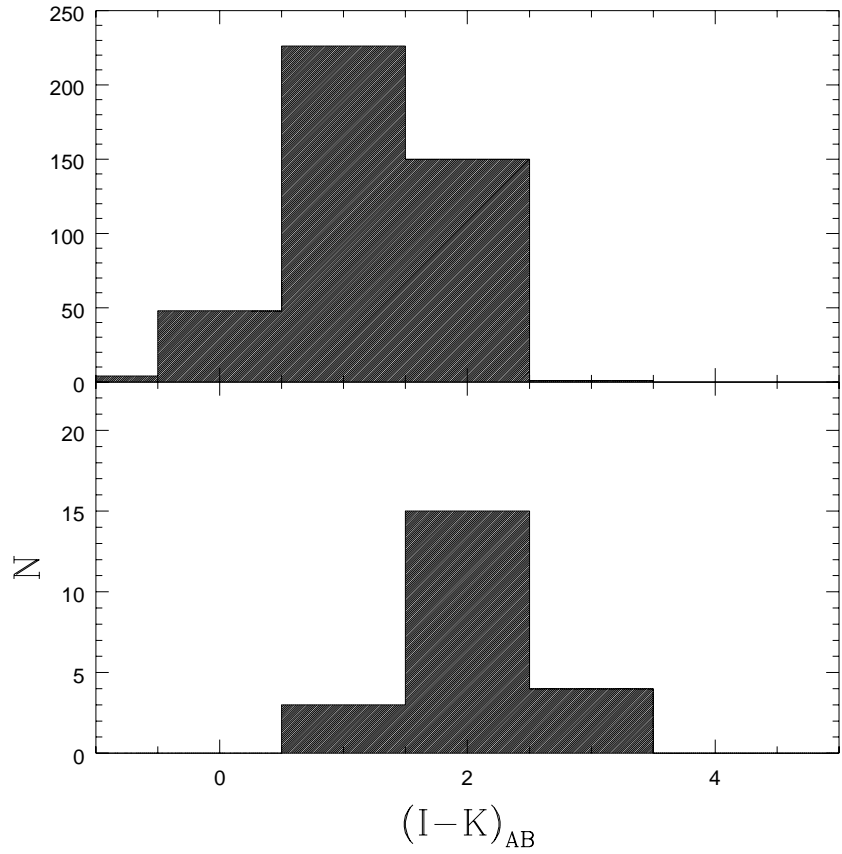,height=12cm,width=12cm}
\caption{ (I-K)$_{AB}$ color distribution of 15$\mu$m sources 
  with $I_{AB} \le$22.5 (bottom panel) 
  compared to the CFRS galaxy color distribution (top panel). }
  \end{center}
\end{figure}

\begin{figure}
\begin{center}
    \leavevmode
    \psfig{file=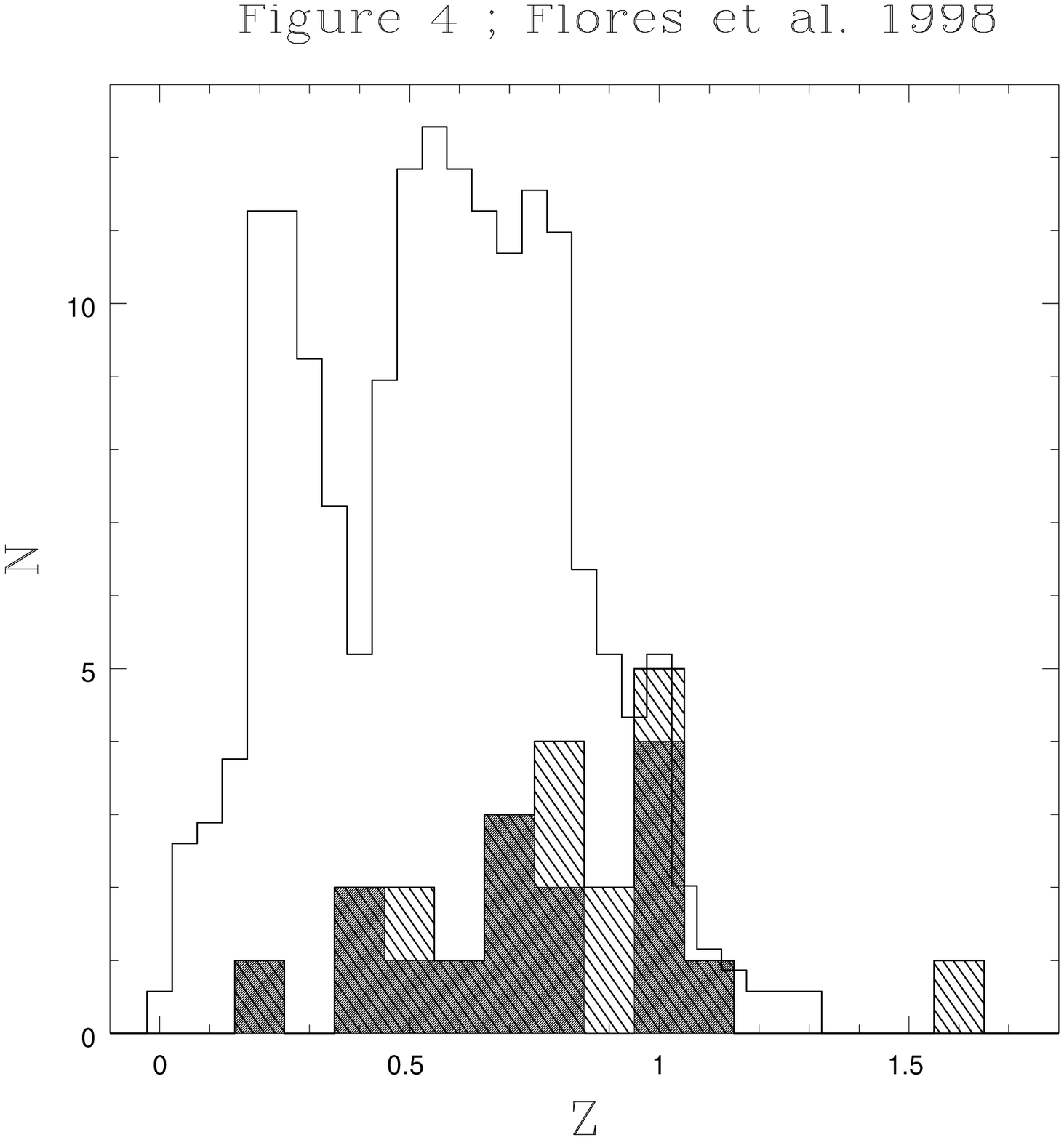,height=12cm,width=12cm}
\caption{ The dashed histogram shows the redshift 
  distribution of the 22 
  identifications with spectra  found in the CFRS database.  The shaded 
  histogram shows the redshift distribution of the sources in the catalogs
 1 \& 2. The non-shaded histogram shows the CFRS redshift distribution after 
  rescaling. }
  \end{center}
\end{figure}

Twenty six of the sources detected at 15$\mu$m and with $I_{AB}\leq 22.5$ have
spectroscopy available in the CFRS database. Among these are 
  4 stars which are also detected at 6.75$\mu$m. Figure 4 shows the redshift
 histogram of ISO sources superposed on the CFRS redshift distribution. The  
median redshift value of the ISOCAM LW3 galaxies ($<z>$ $\sim$ 0.76) is higher 
than that of the CFRS ($<z>$ $\sim$0.58), but coincides with that of
$S_{5GHz}\ge$ 16$\mu$Jy radio sources (Hammer et al. 1995). Since 70\%
 of the 15$\mu$m optical 
sources have $I_{AB}\le$22.5, we estimate from the redshift distribution that 
more than 63\% of the sources with $S_{15\mu m} \ge$ 250$\mu$ Jy are at
 z $\le$ 1.

\subsection{Spectral classifications}
\label{sec:spec_class}

The spectrophotometric classification of the $15\mu$m optical counterparts
is based on optical and emission-line properties  using diagnostic diagrams
and spectral templates (see Hammer et al. 1995, 1997). The following 
spectrophotometric types
 are found: fifteen emission-line galaxies, three active galaxies
 (including the QSO 
CFRS 14.0198 at z=1.6, the most distant CFRS object), one quiescent galaxy, one
 spiral galaxy, and one HII galaxy (CFRS 14.1103).     
Figure 5 shows the spectra of all extragalactic objects, 
classified by their spectrophotometric type.\\

\begin{figure}
\begin{center}
    \leavevmode
    \psfig{file=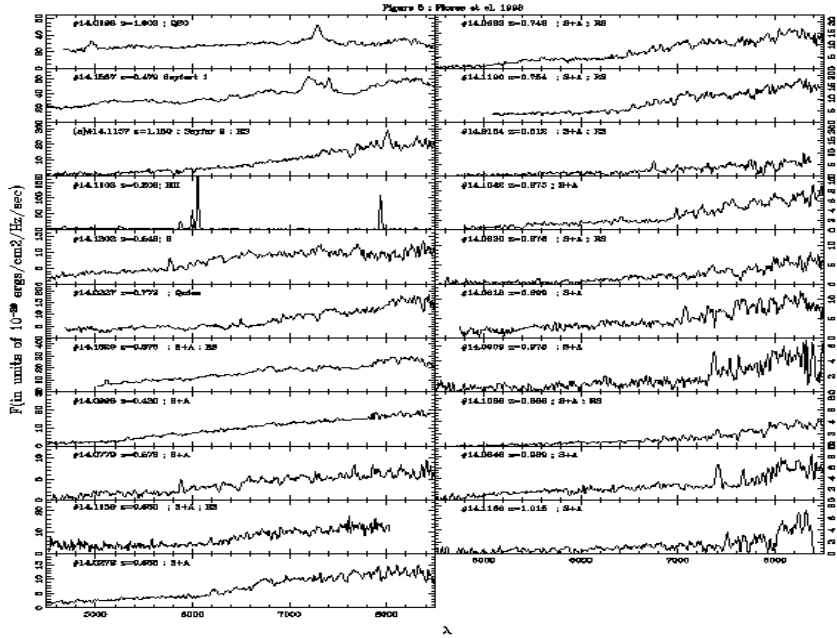,height=12cm,width=12cm}
\caption{ Spectral classification of the 21 15$\mu$m sources with 
available spectra according to their continuum and line properties. For each object, the CFRS 
  name, the redshift and the  spectral classification are given in the upper 
  left corner. }
  \end{center}
\end{figure}
 
Most (71\%) of the 15$\mu$m optical counterparts are classified as  
$S+A$ galaxies, which are galaxies with moderate $[OII]$ emission
($W_{[OII]} \sim$ 20A), and characterized by large D(3550-3850) 
indices (see Table 3), as defined in Hammer {\it et al.}
(1997).
Large D(3550-3850) values can be caused by either a large
 A star population (D(3550-3850) is correlated with the $H\delta$ equivalent 
width, D(3550-3850)=0.2 corresponding to $W(H\delta)$=5-7 A for a non-extincted 
galaxy) or by very large extinctions. As we shall see, it is the A star 
hypothesis which is the most plausible.
This suggests that in most
extragalactic $15\mu$m sources, there were star formation occurring
 about 0.5 Gyr prior to the observed event.  
The importance of the S+A population is also supported by the fact that at
high redshift, galaxies with large D(3550-3850) are also detected at 
6.75$\mu$m and at 1.44GHz.

\subsection{Morphologies from HST and CFHT}
\label{sec:morpho}
About 30\% of the CFRS field at 1415+52 has been observed by the HST. 
Public domain HST images of objects in the
CFRS and Autifib/Low Dispersion Survey Spectrograph survey (the CFRS/LDSS 
deep survey, Brinchmann et al., 1997) and the Groth Survey 
(Groth  {\it et al.}, 1994) are available.
Among the 55 $I\leq 22.5$ ISO sources with optical counterparts,  16 (30\%) 
with have been observed with the HST F814W filter. HST images of
optical counterparts of $\mu$Jy radio sources can be find in Hammer
{\it et al.} (1996).  

\begin{figure}
\begin{center}
    \leavevmode
    \psfig{file=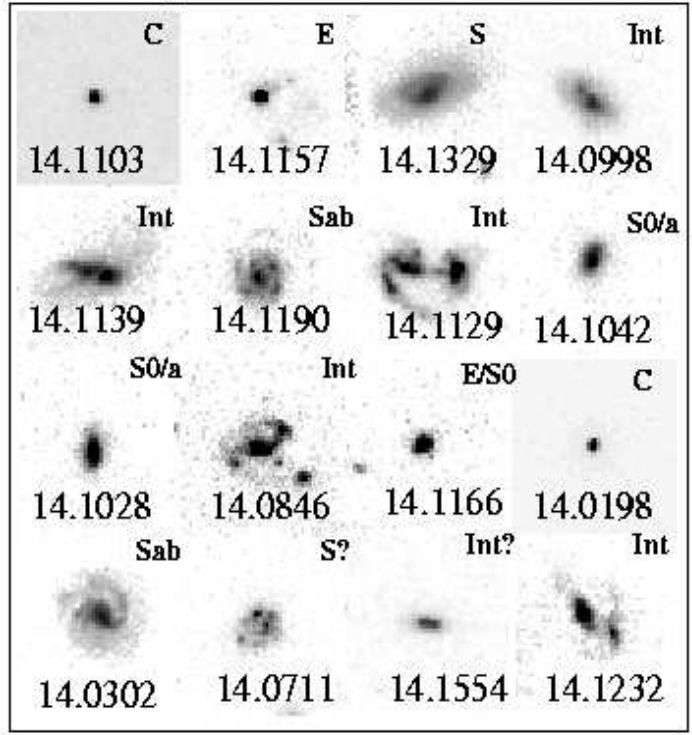,height=12cm,width=12cm}
\caption{ HST 5''$\times$5'' images obtained with the F814W filter
  of 16 sources detected at 15$\mu$m with I$_{AB} < 22.5$. For each
  galaxy, the morphological classification is given in the upper right 
  corner. }
  \end{center}
\end{figure}

Figure 6 displays a mosaic of HST images of optical
counterparts with $I_{AB} \le$ 22.5 of 15$\mu$m sources. 
For each source the CFRS name is indicated as well as 
its morphological classification by either Brinchmann {\it et al.} 
(1997) or us. 

Two sources are unresolved by the HST (one QSO and one HII region, 14.1103).
Six galaxies (37\%) are found in strongly interacting systems, 
four (25\%) are E/S0 galaxies  
and four (25\%) are disk-dominated galaxies. Most of the 14 resolved
sources
show irregularities which might be interpreted as pre- or post-merging events.

\section{Spectral energy distributions and object classifications}
\label{sec:sed}

\subsection{SEDs of local galaxies: toward a template data base}

Several groups have been gathering multi-wavelength observations of
 local galaxies 
(Spinoglio et al. 1995; Schmitt et al. 1998). The main limitations 
of the data sets are the 
assumed aperture corrections and possible AGN variability. SEDs of galaxies
in the same class often show large differences, especially in the infrared.
 This is especially true for Seyfert 2 galaxies which show a wide range of
 colors in the near and far IR (Spinoglio et al. 1995). This large dispersion
 motivated us to 
use mean SEDs of local galaxies to compare with distant sources, so as to
avoid biases for or against a given class of galaxies. For
example, some Seyfert 2 galaxies may be misclassified as starbursts,
 but the effect on global quantities would be compensated by 
true starburst galaxies which are misclassified as Seyfert 2 
galaxies.
However average properties can be affected by a single object with extreme
properties. For example, in the Schmitt et al. (1998) sample,  
the average radio luminosities for elliptical and spiral galaxies 
appears to be overestimated because of the 
large radio powers of NGC 1316 and of NGC 598, respectively. Indeed the mean 
values are  significantly 
($\sim$ 8 times) larger than those derived from the much larger and more complete
 sample of local ellipticals of Wrobel and Heeschen (1991). Using the median 
 rather than the mean 
would provide a value in much closer agreement with Wrobel and Heeschen, and
also provide a SED for spirals which follows the radio-far infrared 
correlation (Condon 1992). In the following we will use only median points
for determining the SED of local templates.\\

\begin{figure}
\begin{center}
    \leavevmode
    \psfig{file=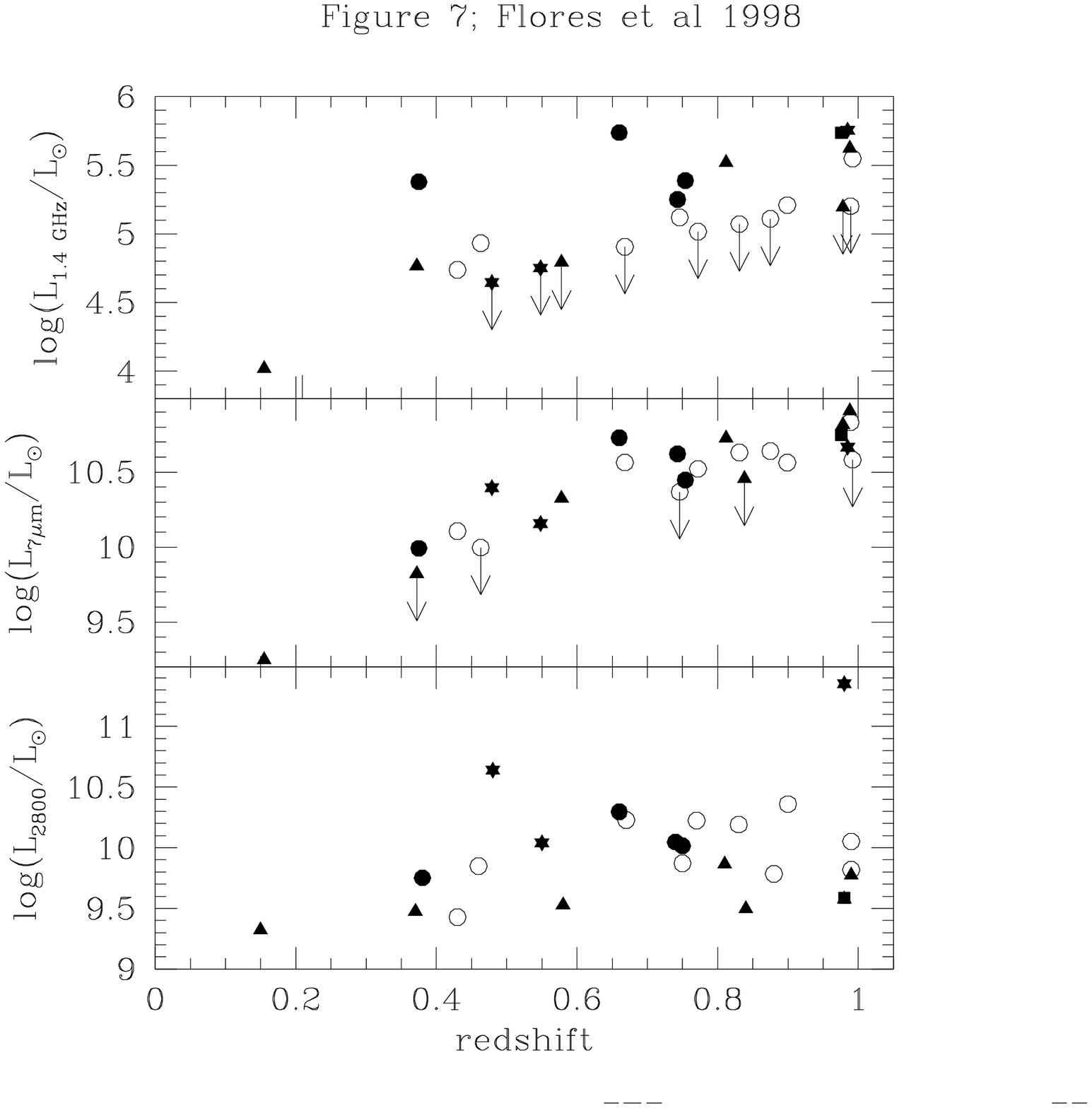,height=12cm,width=12cm}
\caption{ Logarithm of the luminosities at rest 1.4 GHz,
7$\mu$m and 2800A against redshift. Rest-frame luminosities
 have been interpolated
from the observed luminosities at 1.4GHz and 5GHz, at 
0.835, 2.2, 6.75 and 15$\mu$m,
and at 8350A, 5500A and 4350A,
respectively. Filled dots represent SBH, open dots S+SBH, triangles
 Seyfert 2 or Liner and stars QSO or Seyfert 1. Vertical arrows indicate a
detection limit. }
  \end{center}
\end{figure}

 Another concern is whether using local templates to analyze the
  properties of distant sources is appropriate. We have
compared the two samples of local galaxies which have been classified 
and studied at several wavelengths.
 The Schmitt et al. sample has been mainly selected from the IUE 
 archives and the Spinoglio
 et al. sample from the 12$\mu$m faint IRAS catalogue. Our 15 micron sample 
has been selected at the wavelengths 15/(1+z) microns and 0.835/(1+z) microns,
 so that both local samples are
a priori appropriate for use as templates for our sources. 
The two local samples show
somewhat different properties in their SEDs, especially concerning 
the average IR properties of Seyfert 2 galaxies. 
The Schmitt et al. galaxies are bright 
infrared galaxies and include most of the starburst and AGN templates
used by Genzel et al. (1998) to compare with ultraluminous IRAS galaxies.
Conversely, the Spinoglio et al. sample includes galaxies varying over a 4 to 5
magnitude range of luminosities in the infrared, and their average SEDs are
 mostly dominated by fainter galaxies. Figure 7 shows the distribution of 
luminosities for
 the VLA/ISOCAM/CFRS objects. Rest frame luminosities at 7$\mu$ have been
interpolated from broad band observations at 0.835, 2.2, 6.75 and 15$\mu$, and
 are available for most objects, except those detected only at radio
 wavelengths. At z $\ge$ 0.5,these galaxies have luminosities comparable to 
those of the Schmitt et al. galaxies, and are 1.5 to 2.5 magnitudes brighter
than the mean magnitude of the Spinoglio et al. galaxies.  
It is reasonable to think that the discrepancies at IR wavelengths 
between the Spinoglio et al. and Schmitt et al. galaxies are mainly 
due to the large IR luminosity differences.\\
 The large dispersions displayed by individual Schmitt et al. galaxies in each
class (especially for Seyfert 2 galaxies), ensure that this local sample is 
the most appropriate one for comparison with the intrinsically bright distant 
galaxies.

\subsection{Classifications of the 15$\mu$m-radio galaxies from their SEDs}

According to Schmitt et al.(1997), the UV to optical range constitutes
 the best discriminator between Seyfert 2 and SBH or
SBL (starburst with low extinction) galaxies.  Seyfert 2 galaxies
are generally early-type spirals, with UV colors typically bluer than those of
spirals (see e.g. Kennicutt 1992). Figure 8 shows the color distribution of 
the radio-15$\mu$m samples. In the (B-K, 2800\AA~-B) diagram, their colors
are reasonably well fitted by Bruzual and Charlot (1995) models, except 
for three objects which are either a QSO or a Seyfert 1 galaxy. 
They display a range of colors which are well fitted by  
 starburst and spiral templates. The same conclusion can be drawn from the 
  (B-7$\mu$m, 2800\AA~-B) diagram, with a few 
objects having B-7$\mu$ colors consistent with those of Seyfert 2 templates. 
The latter are likely to be associated with hot dust, and 
have redder optical-MIR colors than starbursts. In summary,
most of the galaxies detected at 15$\mu$m have colors from the UV to the MIR
consistent with those of star-forming objects: starburst and spiral galaxies.\\
 
\begin{figure}
\begin{center}
    \leavevmode
    \psfig{file=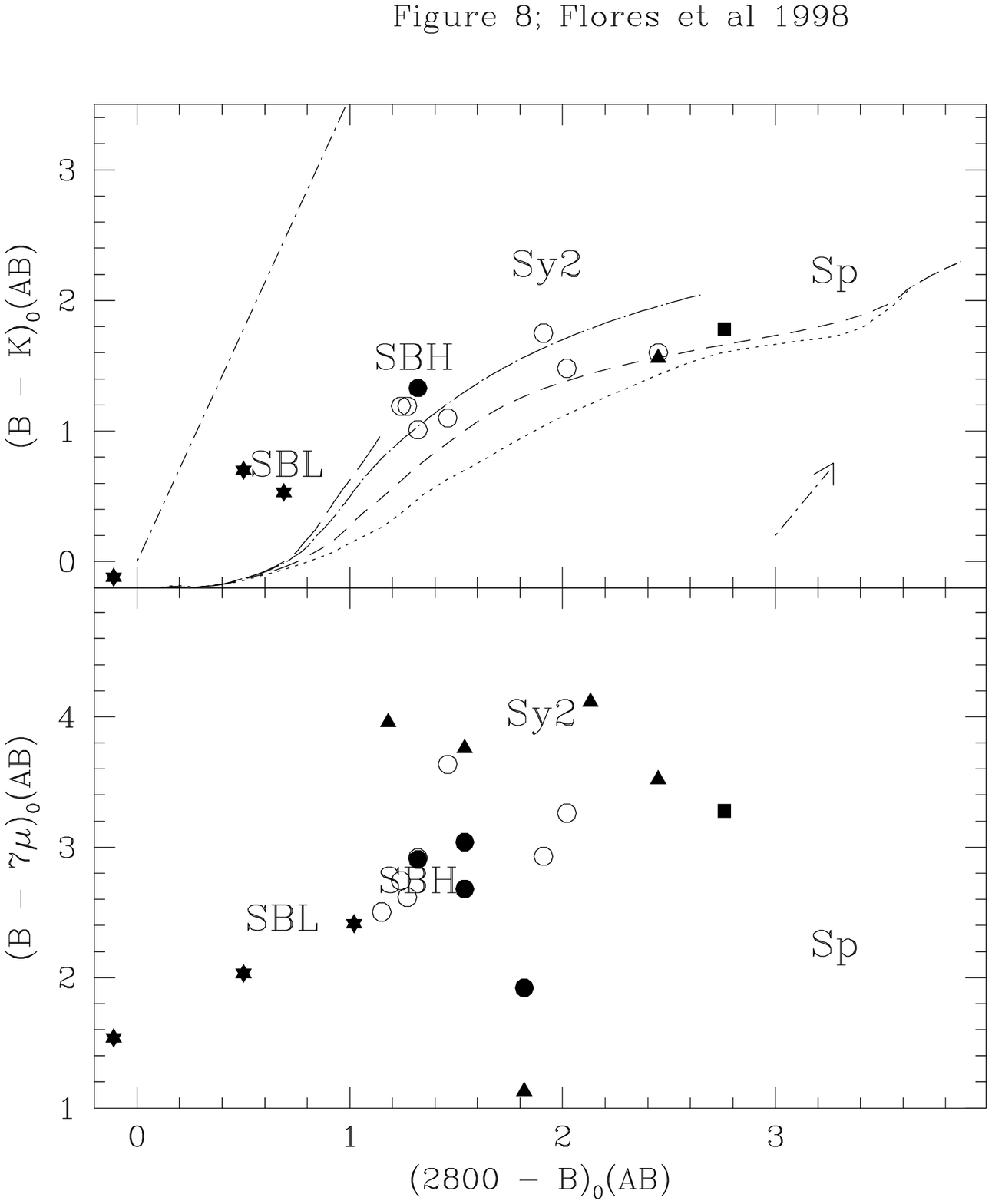,height=12cm,width=12cm}
\caption{ Color-color diagrams for the 15$\mu$m and radio
sources. The symbols have the same meaning as in Figure 7. 
In the top diagram are represented 
galaxies with K band photometry. In the bottom diagram, galaxies
 without infrared measurements are excluded. 
 In the top diagram are shown stellar tracks
 from Bruzual and Charlot (1995, from top to bottom, exponentially decreasing
SFR, with $\tau$=4, 1 and 0.5 Gy, respectively). The dash dotted lines represents
a power law and the extinction vector is indicated. In both diagrams are
shown the color-color location of Schmitt et al. (1998) galaxy templates. }
  \end{center}
\end{figure}

SEDs for the 27 extragalactic 15$\mu$m-radio
 sources with z $\le$ 1 have been constructed, using the observed 
 fluxes at visible
(B$_{AB}$ , V$_{AB}$ and I$_{AB}$ magnitudes), near-IR (K$_{AB}$ when
 available), MIR (6.7$5\mu$m, Flores et al, 1998; 15$\mu$m, Table 1) and
radio wavelengths (1.4 GHz and 5 GHz from Fomalont  {\it et al.}, 1985),
and shifted to rest wavelengths. Each SED has then been  
compared to the average SED of well-known local objects 
(a sample of 59 galaxies, Schmitt  {\it et al.}, 1997). The latter 
include E (ellipticals), Sp (spirals),
SBH (highly reddened starbursts), SBL (low extinction starbursts),
Seyfert 2 and LINERs. Schmitt {\it et al.} (1997) also provided 
standard deviations of the mean, which allows us to optimize the comparison.\\

In order to classify the SEDs, we have performed Monte Carlo simulations,
 assuming independent variations of our flux measurements within 
Gaussian error bars. In each of the 5000 Monte Carlo 
sets, every object SED
 has been fitted to template SEDs spanning all the classes
  defined by Schmitt et al
 , providing a $\chi^{2}$ value
 weighted by the standard deviations of the mean given by those authors.
 Because galaxies
often show properties intermediate between one class and another (see Figure 8),
we also defined a hybrid class which consists of a 
linear superposition of starburst and spiral SED 
templates, with the contribution of the spiral SED varying from
10\% to 90\% in 10\% steps. This hybrid class describes the current  
 mix of old and young stellar populations in a galaxy. We 
have not defined other hybrid classes such as the 
 Seyfert 2 + starburst class, as Schmitt et al. have 
 found that these two populations
 have rather similar radio to infrared properties. 
27 objects were successfully classified with this technique  
(the averaged-weighted $\chi$ is less than or equal to one) except for seven 
(Table 3). Among the latter, four are unambiguously powerful AGN as 
evidenced from their
 optical spectra (see Figure 5 and Hammer et al, 1995), including a QSO
(14.1303) and Seyfert 1 galaxies (14.0573, 14.1302 and 14.1567), and one 
is a HII region (14.1103). 
The 1 sigma dispersion around the $\chi^{2}$ value for each fit 
gives an idea of the reliability of our classification scheme (Figure 9).

\begin{figure}
\begin{center}
    \leavevmode
    \psfig{file=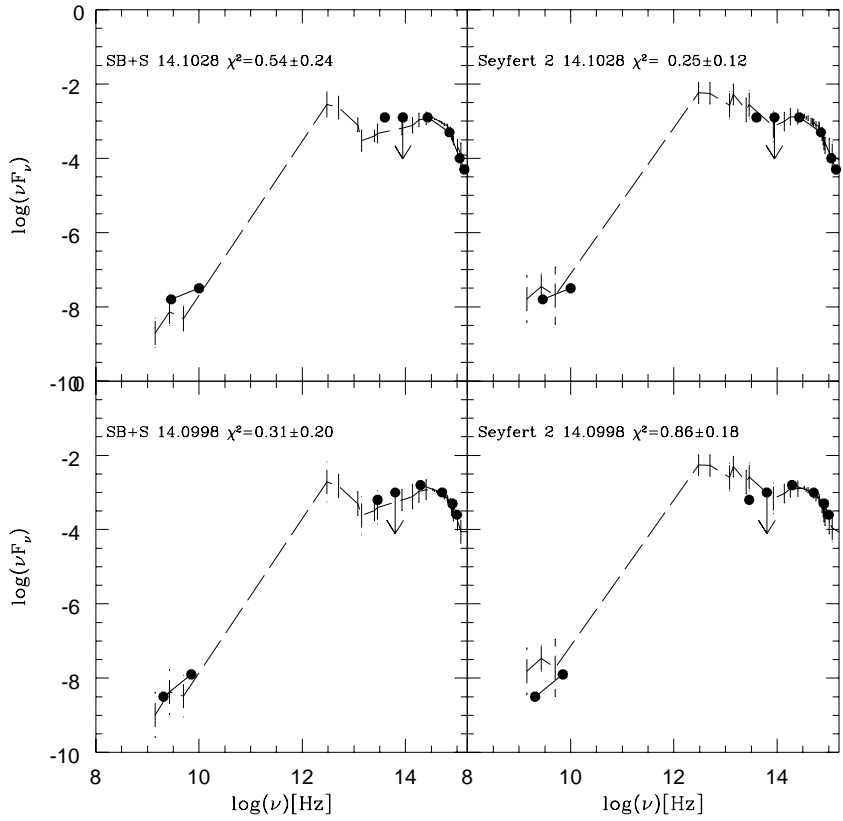,height=12cm,width=12cm}
\caption{ Some examples of possible ambiguous classifications
of the SED.
Solid dots represent flux measurements and are superposed to a template
from Schmitt {\it et al.}. Corresponding $\chi^{2}$ values are indicated (see
text). }
  \end{center}
\end{figure}

\subsection{Final classification of the z $\le$ 1 15$\mu$m-radio sources }

 Among the 27 objects selected either at radio or 15 microns, we find seven 
ambiguous cases, generally in the classes between Seyfert 2 and S + SBH 
(Table 3). Table 4 
summarizes all the informations derived from their optical spectra and
 radio observations (spectral index and 
imagery). For 3 objects, emission lines or radio properties remove the
 ambiguity of the classification: 14.0779 shows a low ionization spectra and 
both 14.1028 and 14.1041 have negative radio spectral indexes and radio
emission extending much beyond their optical sizes).\\
The validity of our classification scheme is illustrated in Figure 8, which 
emphasizes the relevance of the (B-7$\mu$m, 2800A-B) color-color diagram
as a diagnostic diagram
for distinguishing starbursts from AGN. 
In that diagram, starbursts lie in a well defined color-color region,
while AGN-powered sources are around the starburst region. The object
with the smallest (B-7$\mu$m) color is 14.9025 which is classified as a Liner.\\

\begin{figure}
\begin{center}
    \leavevmode
    \psfig{file=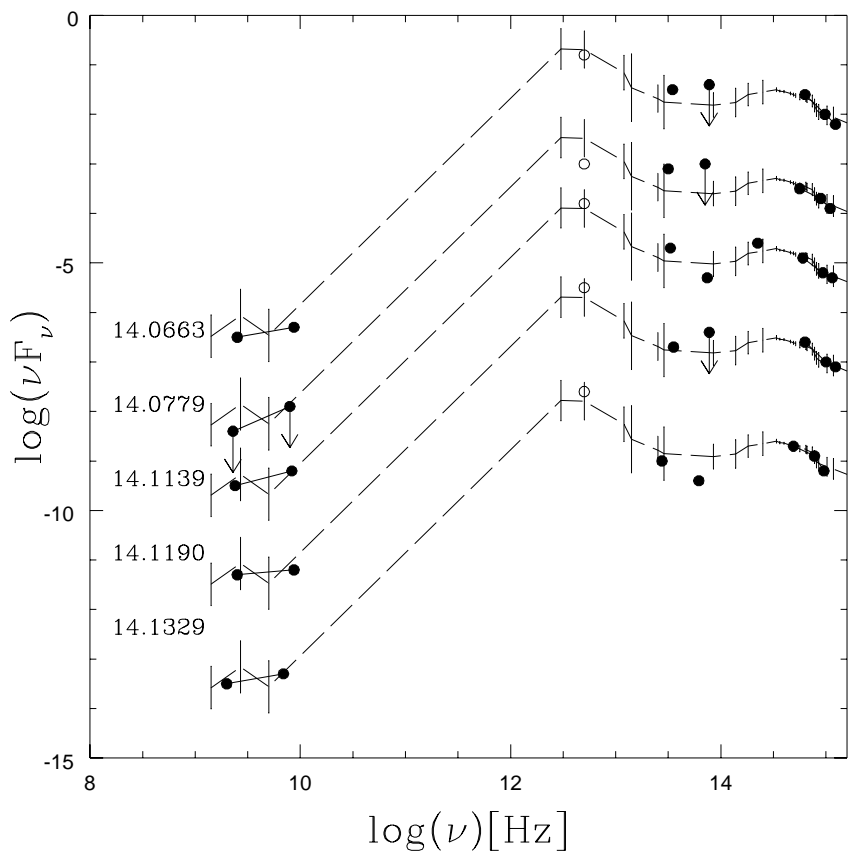,height=12cm,width=12cm}
\caption{ Comparison of the spectral energy distribution 
  (SED) of 5 SBH galaxies (filled circles, separated by arbitrary 
  vertical shifts) with a local 
  averaged SBH SED  (dashed line, from Schmitt  {\it et al.}, 1998). Fluxes 
  at 60$\mu$m (open circles) are derived from radio fluxes according
 to the radio-FIR 
  correlation ($S_{60\mu m}= 125 S_{5Ghz}$, Franceschini  {\it et al.}, 1994).
Superposed to the fit are vertical bars which display at
each wavelength the standard deviation of the local template. The bottom panel
presents the standard deviation of the 5 SBHs. }
  \end{center}
\end{figure}

Five galaxies, all within the 15$\mu$m sample, are classified as pure SBHs. 
They are generally radio sources,
and were classified by Hammer et al. (1995) as being star-forming S+A objects.
Figure 10 displays their SEDs on which is superposed the SBH SED from
Schmitt et al. The interpolated luminosity at 
the 60$\mu$m
 bump and the 60$\mu$m luminosity derived from the radio-FIR correlation
 ($S_{60\mu m}= 125 S_{5Ghz}$, Franceschini  {\it et al.} 1994) agree to 
within 20\% on average.
 Nine objects,
 seven of which are in the 15$\mu$m sample, are classified
 as the superposition
 of a starburst with a spiral SED (Figure 11). For all of them but one 
(14.0846), the starburst component dominates, providing more than 70\% 
of the object bolometric luminosity (see Table 3). Only half of these objects
are detected at radio wavelengths (Figure 7), so they have likely
 lower star formation rates than pure SBHs.\\
It has been widely argued that powerful starbursts such as those 
discovered by IRAS could contain an AGN which can contribute to
their infrared luminosities (Sanders et al, 1988). In addition to the fact that
their energy distributions and colors are typical of those of starbursts,  it is
unlikely that the starbursts described here are significantly 
contaminated by an AGN,
because:\\
-their infrared luminosities ($L_{IR}<$ 2 $10^{12}$ $L_{\odot}$, see Table 4) 
are lower than those of the ultra-luminous IRAS galaxies. Local
galaxies with those luminosities are mainly star-forming galaxies (Lutz et al, 
1998).\\
-their radio spectral indices range from 0.4 to 1, between the observed 
frequencies of 1.4
and 5 GHz. This implies a thermal to non-thermal energy ratio in the rest
frame of 0 to 0.2, in good agreement with the starburst population studied by
Condon (1992).\\
-their radio angular sizes ($<$ 2 arcsec) are always in agreement with their
optical sizes, as expected if star formation was distributed over the galaxy.\\
-five of them have z $\le$ 0.7 and their spectra from CFRS have 
[OIII]5007 and $H\beta$ emission lines; they all show low ionization spectra,
with prominent [OII]3727 line, and for all but one, no [OIII]5007 line. 
Figure 12
shows the sum of the five spectra, which reveals a red spectrum with
$[OII]3727/H\beta$ $\sim$ 1 and $[OIII]5007]/H\beta$ $\le$ 0.05.  These
 galaxies undoubtedly have emission-line spectra typical of HII regions, 
 not of AGN.\\

\begin{figure}
\begin{center}
    \leavevmode
    \psfig{file=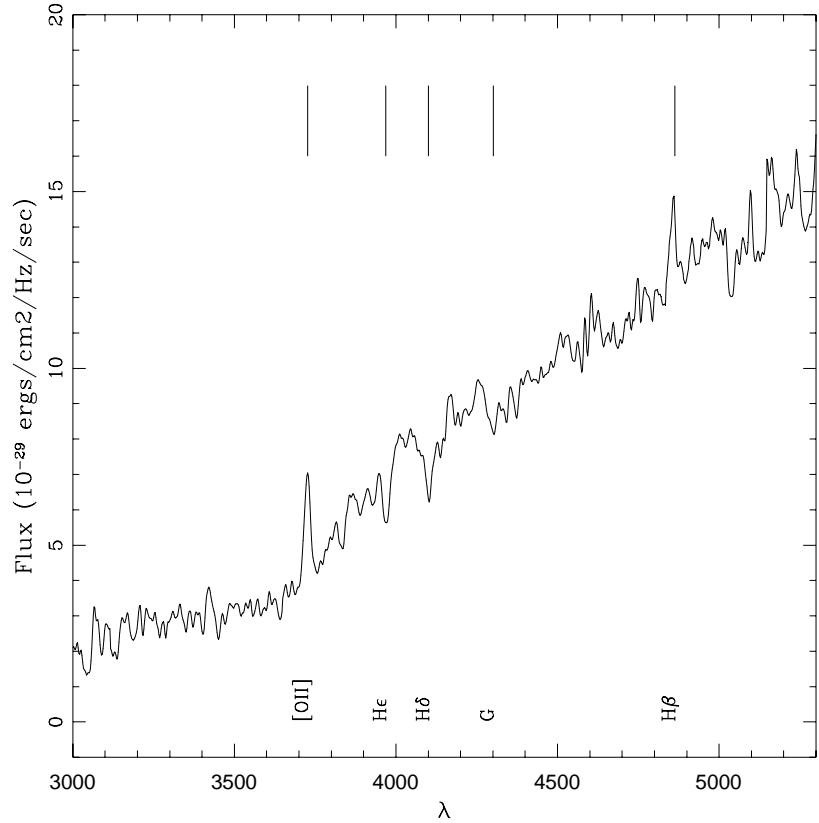,height=12cm,width=12cm}
\caption{ Average spectrum of the 5 SBH and S+SBH galaxies with
z $<$ 0.7. Spectral features are indicated.
 The average spectrum has been computed by using median values
to account for luminosity variation from one object to another. }
  \end{center}
\end{figure}

\begin{figure}
\begin{center}
    \leavevmode
    \psfig{file=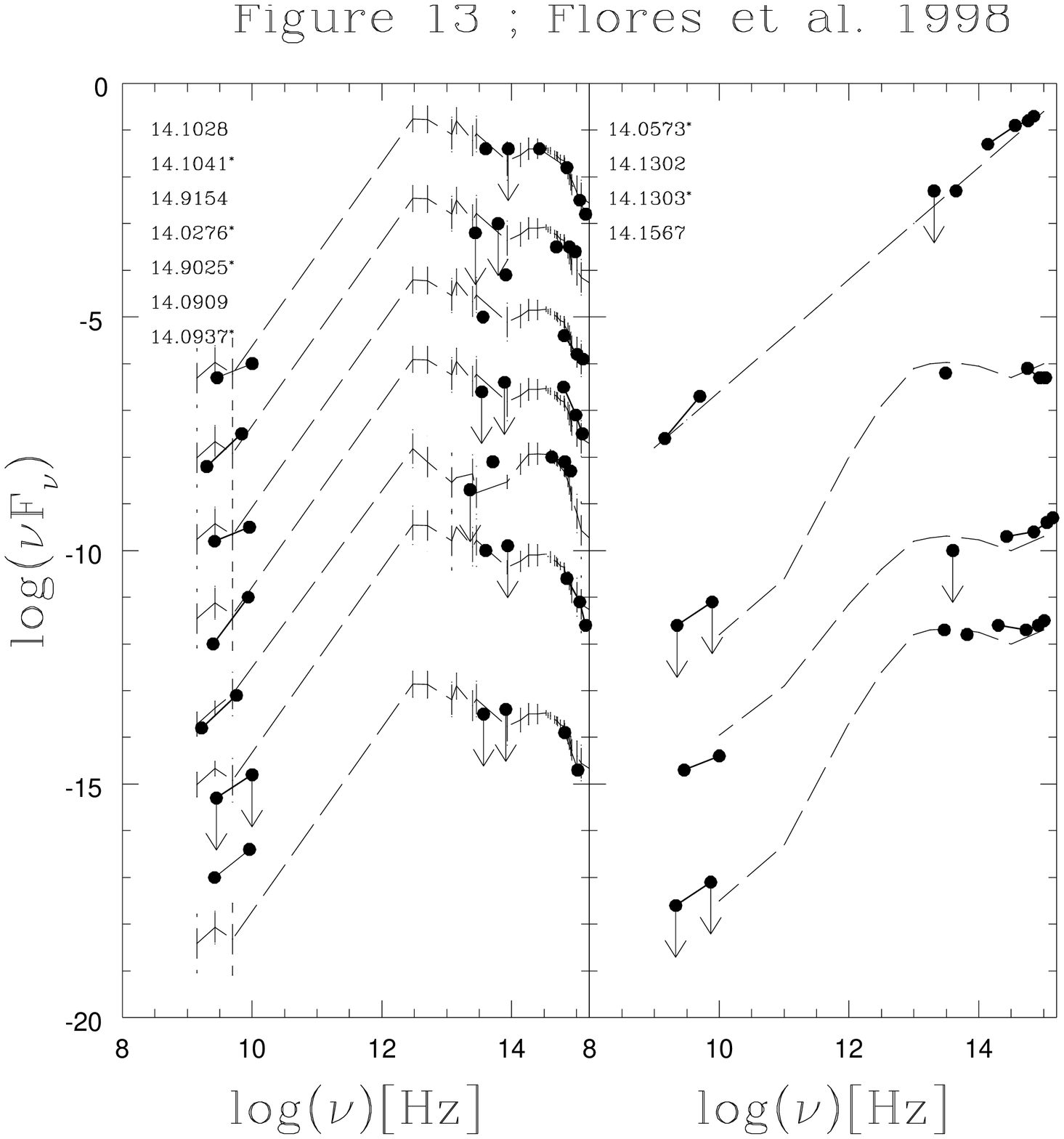,height=12cm,width=12cm}
\caption{Spectral energy distributions (SED) of AGN
  detected with ISO at 15$\mu$m compared with local SEDs. ({\it left}) All
objects are compared with a Seyfert2 SED template except for 14.9025 (Liner
SED template). ({\it right}) Powerful AGN (QSO and Seyfert1) SEDs compared
with power-law or radio quiet QSO (Sanders {\it et al.}, 1989). Galaxies 
  detected only at radio wavelengths are marked with an
  asterisk, and their fluxes at 15$\mu$m by an arrow, assuming a upper 
  limiting 
  flux  of 250 $\mu$Jy.  }
  \end{center}
\end{figure}

Six objects, three of which are in the 15$\mu$m sample, are classified as 
Seyfert 2 galaxies and
 four, two of which are in the 15$\mu$m sample,
are powerful AGN (Seyfert 1 and
QSO). Their SEDs are shown in Figure 13 as well as that of the radio Liner
14.9025, on which are superposed Schmitt et al. templates or radio-quiet
QSO SEDs from Sanders et al (1989).\\
The two remaining objects are 14.1103, an HII galaxy, and 14.0820, an
elliptical or a spiral galaxy with moderate star formation activity
($W_{0}(OII)$=16 \AA).  The object 14.1103 (z=0.21) is detected 
at 15$\mu$m but not at radio wavelengths,
and its SED is consistent with that of a local HII region. 
This is in agreement with the absence of an old stellar 
component as noted by 
Tresse {\it et al.} (1993) from the very large $H_{\alpha}$ equivalent width
($>$ 2500 \AA). A HST image (Figure 6) shows an unresolved object, 
giving a diameter less than 250 pc. We are probably witnessing 
here a compact system undergoing one of its very first bursts of 
star formation: $[OII]_{3727}$ is not detected while
$[OIII]_{5007}$ is very prominent. This galaxy shows spectral properties 
similar to those of a primordial galaxy  
according to Tresse {\it et al.} (1993). Those authors
derived a low heavy-element abundance ($\leq$ 0.05 the solar value) and a large
effective temperature (T$>$50 000K based on the 
$[OIII]_{4363}$/$[OIII]_{5007}$ ratio).\\

\section{Global UV and IR luminosities}
\label{sec:sfh}

\subsection{Global UV luminosity}

$L_{2800}$ values have been interpolated from $L_{4350}$ and $(V - I)_{AB}$
(or $(B - I)_{AB}$) colors, using a grid of models with 
exponentially decreasing SFR ($\tau$=1 Gyr, Bruzual and Charlot, 
1995) and a Salpeter IMF (dN(m)/dm=-2.35 over m=0.1-100 $M_{\odot}$). These
calculations are found to be rather independent of the exact value of $\tau$, 
except for very small values ($\tau <<$ 0.1 Gyr). We have calculated the total 
L$_{2800}$ (see Table 5) in the CFRS 1415+52 field. It is based on 143 galaxies
which have been spectroscopically identified and are representative of the 
558 $I< 22.5$ galaxies in the field (see Lilly et al, 1995b). Converted to a  
luminosity density, this value is in excellent agreement (to within 6\%)
 with that found for the whole CFRS field (Lilly et al, 1996).
This is consistent 
with the Poisson error of 8\% for the 143 representative galaxies.   
 
\subsection{Global infrared luminosity}

\subsubsection{Interpolated infrared luminosities}

Infrared (8-1000$\mu$m) luminosities are interpolated from our MIR and radio
flux measurements, by using for each object the template from
Schmitt et al. (1997) with  
the smallest $\chi^{2}$. The template gives 5 to 7 points 
within the wavelength range of interest, 
and errors can be estimated for each object using our Monte Carlo
simulations. Our quoted errors take into account flux measurements errors 
(13\% on average)
as well as uncertainties in our classification scheme (5\% on average).
The interpolated infrared luminosity for each object
along with its associated error is given in Table 4.   

\subsubsection{Global infrared luminosity from galaxies of the 15$\mu$m and 
radio samples}

Galaxies detected at 15$\mu$m and/or at radio wavelengths have infrared 
luminosities in the range of 5 $10^{10}$ to 2 $10^{12}$ $L_{\odot}$ (Table 4),
 comparable to those of typical local starbursts and Seyfert 2 galaxies 
(Genzel et al. 1998). These should be compared to their 2800\AA\ luminosities 
which range from  4 $10^{8}$ to 2 $10^{10}$ $L_{\odot}$ and 
are not very different from those of galaxies not detected at 15$\mu$m 
and at radio wavelengths. Thus, taking into account only the UV luminosity
of the 15$\mu$m and radio galaxies which are luminous 
infrared sources would lead to severe underestimates 
of their actual star formation rates. Nearly half of the global infrared 
luminosity of 15$\mu$m and radio galaxies is coming from 7 sources 
(4 SBHs and 3 Seyfert 2). These are detected at both wavelengths,
except for the Seyfert 2 galaxy 14.0937. 
Estimates of infrared luminosities of these bright galaxies should thus  
not be affected strongly by source count biases or modelling 
uncertainties.\\

Accounting for the global luminosity is slightly complicated by the 
current observational
 status of the data. Hammer et al. (1995) and recent follow-up work have nearly 
completed the redshift identifications of the radio $\mu$Jy sources. Among 
the 45 $I_{AB}\le$ 22.5 15$\mu$m sources which are not detected at radio 
wavelengths, only 16 possess a redshift from the CFRS. 
The latter have not been selected {\it a priori} for their infrared 
properties, and can be considered as representative of the whole sample. 
In the calculation of 
the global luminosity, we have assumed two different scaling factors:
1 for luminous galaxies detected at radio wavelengths, and 45/16 = 2.8 
for the less luminous galaxies undetected in radio, with an additional 
statistical error of 25\%. Table 5 presents the different values of the 
global luminosities at 2800\AA\ and in the infrared,
 after excluding powerful AGN 
(QSO and Seyfert 1).  At IR wavelengths, Seyfert 2 
contribute to more than a third of the global luminosity.

\subsubsection{Global infrared luminosity from galaxies undetected at 
15$\mu$m and at radio}

We consider here the subsample of the 49 galaxies which possess a redshift,
are not detected at 15 $\mu$m or radio wavelengths,
and have been observed at K. This subsample can be taken as 
representative of the 489 $I_{AB} \le$ 22.5 galaxies not detected
at 15$\mu$m or at radio wavelengths in the CFRS 1515+52 field. 
For these galaxies, we use BVIK photometry and their 15 $\mu$m and radio 
upper limits to
investigate their SEDs. We found that 19 of them have SEDs consistent
with those of local template SEDs for SBH galaxies and 16 with local 
template SEDs for SBL galaxies. The relatively large fraction of SBLs 
(starbursts with low extinction) is compatible with the fact that they 
are not detected in the MIR range. Other objects have 
their SEDs consistent with those of local template SEDs for E/S galaxies 
(9 objects) or AGN galaxies (5 objects). The fraction of AGN is in reasonable
agreement with Hammer et al. (1997) who find that 
 $\sim$ 8\% of all galaxies in the field  
 at z $\sim$ 0.5, are Seyfert 2 galaxies as indicated by their emission 
 line ratios.   
Figure 14 displays four examples of fits
for these objects. All objects but four have been reasonably classified by
this method. The 4 remaining objects have been finally classified as starbursts.\\

\begin{figure}
\begin{center}
    \leavevmode
    \psfig{file=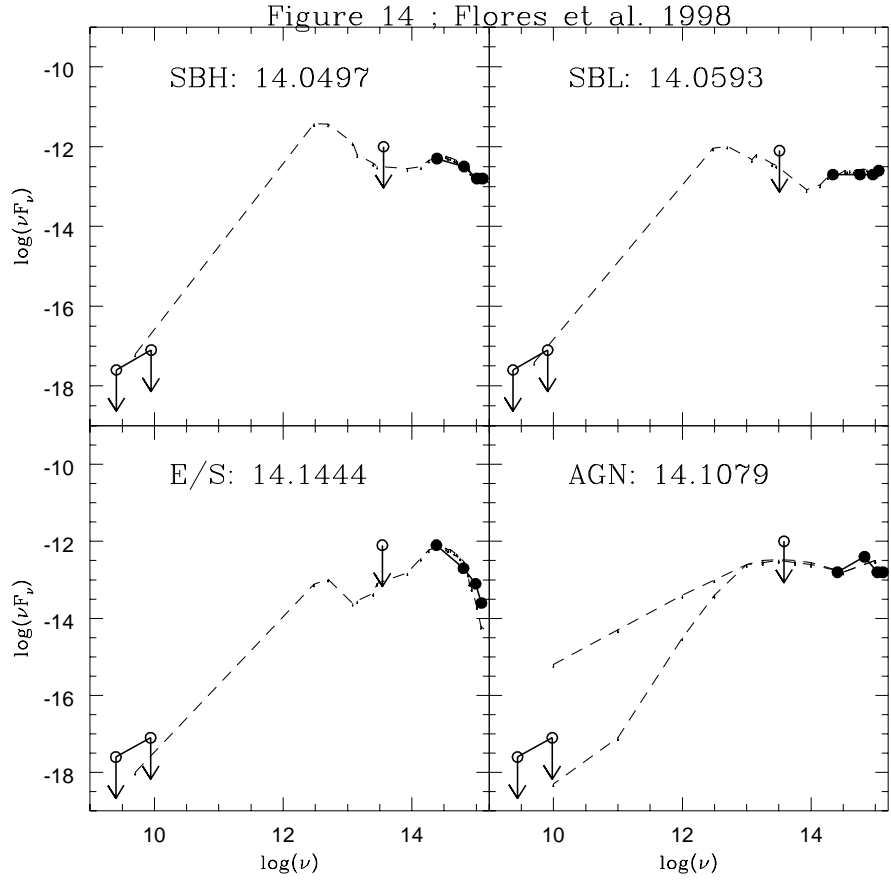,height=12cm,width=12cm}
\caption{ Four examples of the spectral energy distribution of 
galaxies which are not detected at 15$\mu$m and at radio wavelengths, but 
  possess B to K observations. }
  \end{center}
\end{figure}

\subsubsection{Infrared luminosity density related to star 
formation and uncertainties}

The luminosity density of star-forming galaxies may be simply estimated
by summing up the contribution of all starburst galaxies (SBH, SBL and S+SBH)
identified in the field. Uncertainties which include errors in source counts, 
flux errors and uncertainties in the identification, would come from  
our Monte-Carlo simulations.\\ 

However several complications could affect the
validity of such a calculation. First, the origin of the UV and
infrared light from Seyfert 2 galaxies is unclear, and can be
attributed either to the AGN or to star formation or to both.
 Indeed Seyfert 2 galaxies show
a wide range in their infrared properties, and few color diagrams
are able to distinguish them from starbursts. Contamination 
of the luminosity density by AGN is potentially the most serious problem
for any estimate of the star formation density. Second,
 cirrus can contribute to the infrared luminosity, and
this is likely the case for the objects classified as S+SBH. 
It is often assumed that the 60$\mu$m luminosity of apparently quiescent 
early-type galaxies is related to cirrus excited by the emission of the 
underlying stellar population (Sauvage \& Thuan, 1992).
 This is however disputed by Devereux and Hameed 
(1997) who present some counter-examples, 
and argued that young stars can be an
 important source of the FIR light of early-type galaxies. One
may wonder however about the exact nature of early-type galaxies 
with a significant
population of young stars embedded in dust. Might they not be equivalent to our
S+SBH galaxies? In that case, the cirrus contribution to 
the infrared luminosity of these galaxies 
cannot be significant as the starburst 
component contribute to most of their bolometric luminosities.\\ 
    
Recall that former estimates of the SFR co-moving density from the UV light are 
uncertain, because:
\begin{enumerate}
\item they include contributions at 2800\AA\ from AGN and old 
stellar populations.
\item they do not account for the UV light reprocessed by dust into FIR 
radiation.
\end{enumerate}

We wish here to take into account the deficiencies mentioned above and 
calculate {\it upper and lower limits to the luminosity density related to 
star formation}. The upper limit is obtained with the following assumptions:\\

First, the 15$\mu$m and radio fluxes of all objects not detected at those 
  wavelengths are set equal to their detection limit. This provides 
an upper limit to their infrared flux.\\

Second, the 2800\AA\ and 60$\mu$m luminosities of AGN-classified objects are 
  assumed to be powered by star formation.\\

Third, the infrared luminosities of S+SBH galaxies are assumed to be only
coming from star formation.\\

The lower limit is obtained with  the following assumptions:\\

First, all objects not detected at 15$\mu$m and at radio wavelengths have 
  zero extinction, and hence the star formation rate estimated
 at infrared wavelengths equals that estimated at UV wavelengths.\\

Second, the 2800\AA ~and 60$\mu$m luminosities of AGN and of 
  quiescent (E) galaxies are assumed to be related respectively to the 
  active nucleus and to the old stellar component.\\

Third, the infrared luminosities of S+SBH galaxies are assumed to be related
to both cirrus and star formation, in the proportion of 
 the ratio of the bolometric luminosities of the two components.\\

Table 5 displays the global luminosities in the CFRS 1415+52 field, 
as derived from UV 
(2800\AA ~) and infrared (8-1000$\mu$m) fluxes, 
for the sample of objects detected either
at 15$\mu$m or at radio wavelengths, and for that containing objects not 
detected at both wavelengths. The AGN contributions are given in the upper 
limit case. They can attain non negligible values ($\sim$ 20\%), 
including at UV wavelengths. This represents a major uncertainty
for the SFR densities derived from UV fluxes. 
The errors take into account uncertainties in source counts,
 flux measurements errors and uncertainties in
the classification scheme.

\section{Estimation of the cosmic star formation rate}

\subsection{Star formation density derived from the UV light density}

The star formation history has been estimated by various 
 authors (see  Madau {\it et al.} 1996; Lilly {\it et al.} 1996; 
Hammer {\it et al.} 1997) 
on the basis of the CFRS and/or the HDF surveys, using star formation 
rates derived from 2800\AA\ or $[OII]_{3727}$ fluxes. 
Since the UV light is dominated by emission from more or less massive stars,
the total SFR as derived only from the flux at  2800\AA ~ is somewhat uncertain,
as the extrapolation to the low mass end ($M \leq 5 M_{\odot}$) of 
the IMF is not constrained. Moreover
Hammer {\it et al.} (1997) have suggested that dust and metallicity can
severely affect these estimates and mask the true evolution of the cosmic
star formation rate.\\
We adopt in the calculation of
the star formation rate (SFR) from the UV light the calibration
of Madau et al (1998) and Kennicutt (1998), assuming a Salpeter (1955) IMF with
mass limits 0.1 and 100 $M_{\odot}$:

\begin{equation}
SFR_{2800}= 5.045 X 10^{-10} (L_{2800}/L_{\odot})  
\end{equation}

\subsection{Star formation density derived from the infrared light density}

The SFR can also be inferred from infrared luminosities, assuming that they 
result mostly from dust heating by young stars, in the optically thick limit.
 The SFR calibration depends mainly on the
burst duration and on the IMF slope at both high and intermediate mass ranges.
Following Kennicutt (1998) we adopt the same IMF as that assumed to calculate
the SFR derived from UV light and:\\

\begin{equation}
SFR_{IR}= 1.71 X 10^{-10} (L_{IR}/L_{\odot})  
\end{equation}

The above equation is based on the models of Leitherer \& Heckman (1995) 
for a continuous burst with age 10-100 Myr. 
Condon (1992) calibrated the SFR from the non-thermal 
radio luminosity ($L_{NT}$), 
assuming the Galactic relation between $L_{NT}$ and the 
radio supernova rate. He then derived the $SFR_{IR}$ calibration from the 
FIR-radio correlation, showing that $\sim$ 2/3 of the 
UV-optical emission is re-emitted between 40 and 120 microns 
(see also Helou 1988). After rescaling his value by assuming a Salpeter IMF 
(0.1-100 $M_{\odot}$), Condon (1992) obtains:

\begin{equation}
SFR_{IR}= 2.22 X 10^{-10} (L_{IR}/L_{\odot})
\end{equation}

The difference between the two authors is at the 28\% level, in agreement
with the Kennicutt uncertainty estimate of $\pm$ 30\% on his calibration, 
after comparison with other models (Hunter et al, 1986, Lehnert \& Heckman,
1996, Meurer et al, 1997).\\

The key parameter in these estimations is the star formation rate of massive
 stars (typically $\ge$ 5 $M_{\odot}$), which are mainly responsible
for the UV continuum, the non-thermal radio continuum and FIR 
luminosities re-radiated by dust. The $SFR_{IR}/SFR_{UV}$ ratio
calculated in the following, is essentially independent of the IMF slope,
 and in particular of any 
extrapolation towards low stellar masses.

\subsection{Star formation density missed by UV observations}

 We aim to calculate here  the cosmic SFR from the infrared emission, 
and  compare it with that derived 
from the UV emission. We first consider the "directly observed"
SFR,  based on observations of the 
$I_{AB} \le$ 22.5 galaxies, as presented by Lilly {\it et al.} (1996). 
In section 5.1, we have calculated the total $L_{2800}$ 
contributed  by all galaxies with $I_{AB} \le$ 22.5 and $z \leq 1$. 
In section 5.2, we have computed the total $L_{IR}$
from those galaxies with either $S_{15\mu m} \ge$ 250 $\mu$Jy or 
$S_{5GHz} \ge$ 16 $\mu$Jy and from CFRS galaxies not 
detected at 15$\mu$m and at radio wavelengths.\\

We have calculated the $SFR_{IR}/SFR_{2800}$ ratio accounting for all galaxies 
not classified as powerful AGN. It ranges from 2.5 $\pm$0.95 in the lower 
limit case, to 5.4 $\pm$ 1.9 when galaxies undetected at 15$\mu$m and 
in radio have their 
fluxes at these wavelengths equal to the detection limit. These values can 
however be affected by uncertainties on the calibration of the SFR in the  
UV and FIR wavelength ranges. Nevertheless the values we find for 
$SFR_{2800}$ and $SFR_{IR}$ show some consistency, since for all galaxies,
 $SFR_{IR}$ is larger than $SFR_{2800}$.  These ratios 
give extinctions ranging from $A_{V}$=0.49 to 0.87, if we assume a standard 
galactic extinction law. If the true SFR density is given 
by the FIR estimation, then 
from 35\% to 85\% of the global SFR co-moving density for z $\leq$ 1 
are not taken into account when only the UV flux density is considered.\\
\\
    
Figure 15 shows the co-moving SFR density evolution with lookback time
(for $H_{0}$=50 km $s^{-1} Mpc^{-3}$ and $q_{0}$=0.5). The VLA-ISO-CFRS
points (filled dots) represent the average values of the upper and lower 
limits defined above. Since there is no evidence for a change in the 
extinction for z $\leq$ 0.5 as compared to for 0.5 $<$ z $\leq$ 1,
 we have adopted the same redshift bins as those used in previously
 deriving the SFR from the UV fluxes. VLA-ISO-CFRS data are 2.9$\pm$1.3
larger than former UV estimated values, even though latter were accounting for
all sources, including strong AGN. The error bars are accounting for 
incompleteness errors (Lilly et al, 1996) as well as uncertainties in 
the determination of the SFR density at IR wavelengths. \\

\begin{figure}
\begin{center}
    \leavevmode
    \psfig{file=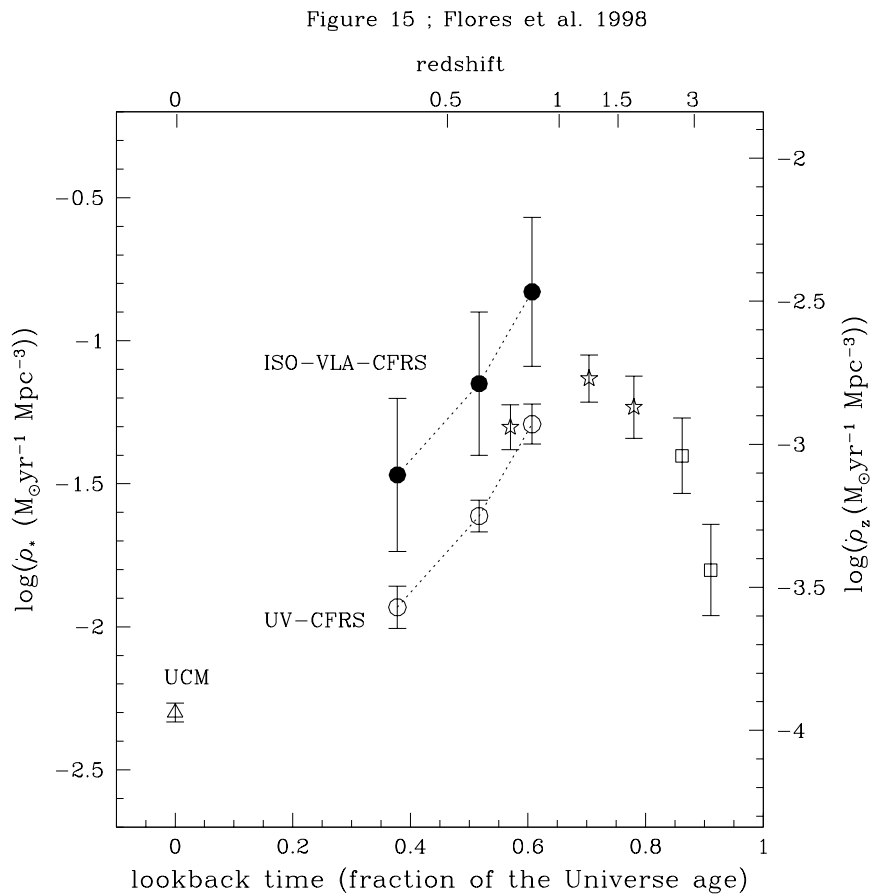,height=12cm,width=12cm}
\caption{ Metal production and star formation history for 
$z\leq 1$ (see text). Our
  points (filled circles, labeled ISO-VLA-CFRS) are 2.9
   times higher 
in SFR density or in metal production than those 
(open circles) previously derived from 
the UV flux density at 2800\AA. Other points are 
  from Gallego {\it et al.} (1995, open triangle), Connolly{\it et al}
(1997) (open stars), Madau {\it et al.} 
(1998, HDF) (open squares), and have not been corrected for extinction. }
  \end{center}
\end{figure}

Error bars in Figure 15 are very large mostly because we have assumed
that the blue galaxies undetected at 15$\mu$m and at radio wavelengths
have no extinction (lower limit case), or have flux densities
equal to the detection limits
(upper limit case), both of which are clearly unrealistic assumptions.
It is interesting to check whether our average value makes sense.     
Starbursts detected by both ISOCAM and the VLA are rare. They have large 
extinctions ranging from
 $A_{V}$= 1.5 to 2, and show red continua (Figure 12). Tresse \& Maddox
 (1998) predict
 an extinction of 1 mag at 2800\AA~ (or $A_{V}$= 0.48) for the blue field
 galaxy population
 at z $\le$ 0.3. If the same value holds on average for blue galaxies
 up to z = 1, 
we can predict in principle the infrared luminosity of the field galaxy 
population not detected at 15$\mu$m and at radio wavelengths. For them, 
Table 5 gives an
 average value $L_{2800}$= 196 $10^{10}$ $L_{\odot}$ in the field. 
According to equation 3, this 
 corresponds to a global $SFR_{2800}$ of 980 $M_{\odot}yr^{-1}$. 
Adopting the Tresse \& Maddox extinction estimate, the global infrared
luminosity would be equal to 1445 $10^{10}$ $L_{\odot}$ (equation 4). This is 
in very good agreement with the average value in the fourth column of 
Table 5, and adds more confidence in the VLA-ISO-CFRS average values shown 
in Figure 15.\\
    
 These average values imply that the stellar mass formed since z = 1 would be 
3.3 $ \times 10^{8}$ $M_{\odot} Mpc^{-3}$, slightly larger than the estimation
of the present day stellar mass of 3 $ \times 10^{8}$ $M_{\odot}Mpc^{-3}$ 
(Cowie et al., 1996; Glazebrook et al., 1995). 
Of course, the uncertainties on 
the SFR estimations are large,
 and our lower estimate is still consistent with the
present day stellar mass. However we believe that this will 
constitute a serious problem,
when uncertainties will be lowered by larger and deeper surveys at infrared
 and radio
wavelengths. The problem gets worse if one accounts for the additional star
formation at z$>$ 1. There may be two ways out.
First, half of the SFR density comes from $\sim$ 10\% of the field galaxies,
 those detected
at 15$\mu$m and at radio wavelengths. These objects
 with large SFR (especially those with $>$ 100 
$M_{\odot}$$yr^{-1}$) may have IMFs which significantly deviate from the
 Salpeter law. Alternatively, the local stellar mass density may have been
systematically underestimated.

\section{Comparison with the HDF/ISOCAM surveys}

The number densities of faint MIR and radio sources found in the 
CFRS 1415+52 field and in the HDF appear to be a reasonable agreement. 
However our results are not consistent with those derived for the HDF
 by Rowan Robinson {\it et al.} (1997). The HDF is 18.5 smaller in area than
the CFRS 1415+52 field, and from our results, we would expect 0.3
starburst with SFR larger than 100 $M_{\odot} yr^{-1}$, as compared 
to the four starbursts with SFRs from 500 to 1010 $M_{\odot} yr^{-1}$
 claimed by Rowan-Robinson {\it et al.} (1997). No simple explanation can 
account for this discrepancy of more than one order of magnitude.
The calibration of the $SFR_{FIR}$ by Rowan-Robinson et al(1997) differs 
slightly from ours (their proportionality factor in equation 4 is 
2.6 $10^{-10}$ instead of our 1.7 $10^{-10}$). We simply stress that:
\begin{description}
\item[a)] our source detection and identification procedures are robust 
  (see also Flores et al, 1998), 
  while source identifications can be difficult in 
  the HDF, since several galaxies in the HDF can lie within a
 single ISOCAM pixel.
\item[b)] Rowan Robinson {\it et al.} (1997) have used a three-component model 
  (a stellar component giving rise to the UV to MIR emission plus
  cirrus and starburst components giving rise to the MIR to radio emission)
 to fit galaxy SEDs. This type of  model has several degrees of freedom, 
  including the relative energy ratio between the different components. The 
  number of free parameters is larger than the number of available data points
(1-3) in the MIR to radio wavelength range.
\item[c)] Among the twelve HDF galaxies whose SEDs were fitted by 
Rowan Robinson {\it et al.} (1997), five have 
  limits on their radio luminosities which are significantly lower than 
the starburst model predictions (see their Figure 1).
\end{description}
Instead  of force-fitting the data points by multiple-parameter starburst 
models, we propose here a more conservative empirical approach of fitting SEDs
 of galaxies with $z \leq 1$ with SED templates of well-studied 
local galaxies (see Schmitt et al. 1997). The fits are generally quite good, 
especially for starburst (SB) SEDs which  are consistent with the standard
radio-FIR correlation. A pure starburst has its energy distribution dominated
 mostly by star formation. Its SED includes massive star light with or without
extinction at UV wavelengths, reemission of the absorbed UV light at MIR 
and FIR wavelengths, and starburst thermal emission and synchrotron radiation
from supernova remnants at radio wavelengths. Locally, 
Schmitt {\it et al.} (1997) 
show that starbursts, when compared to normal spirals or ellipticals,
 have a smaller spread in their SEDs from FIR to radio
wavelengths. There is no evidence that these properties change 
significantly in the redshift range 0 $\leq  z \leq 1$.

Strong starbursts (with SFRs larger than 70 $M_{\odot} yr^{-1}$)
should be detected by deep radio surveys to $\mu$Jy levels, such as those by
Fomalont {\it et al.} (1992) in the CFRS 1415+52 field,
 or by Fomalont {\it et al.} (1997)
 in the HDF. The fact that these are not detected in the HDF at radio 
wavelengths casts some doubts on the large SFRs derived by Rowan Robinson 
et al. (1997) for objects without radio counterparts.         
As for the 3 radio sources with $S_{5GHz}\ge$ 16$\mu$Jy and z $\le$ 1 which are 
found in the HDF (Fomalont et al, 1997), one (ISOHDF 12 36 46 + 62 14 06, 
z=0.96) is found by Rowan-Robinson {\it et al.} (1997) to have the
highest SFR (1010 $M_{\odot} yr^{-1}$) in their sample.
It is at z=0.96, has an elliptical morphology, is undetected at 15 $\mu$m
(as most of the elliptical galaxies of the Hammer {\it et al.} (1995)
 radio sample), and shows highly ionized emission typical of a Seyfert galaxy 
(MgII$\lambda$2799, [NeV]$\lambda$3426 and [NeIII]$\lambda$3868).
It is likely a Seyfert 1 galaxy since the MgII line is broad, and 
its radio emission shows some variability (Richards et al. 1998). 
It is doubtful that its emission is powered by star formation.\\

To derive FIR luminosities (and hence SFRs) requires complete information
from UV to radio wavelengths. It seems reasonable to use templates
based on well-known local objects, especially for the most powerful
star-forming galaxies.   

\section{Conclusions}
\label{sec:discussion}
Observations of distant field galaxies have been obtained with ISOCAM down
to unprecedented flux levels at 6.75$\mu$m (S$\ge$150$\mu$Jy) and at
15$\mu$m (S$\ge$250$\mu$Jy). We have attempted to make our samples at both 
those wavelengths as complete as possible, by a  careful reduction of the 
data. That the samples are complete is suggested by the good positional 
correlation between source identifications at 6.75$\mu$m,
15$\mu$m and radio wavelengths.\\
\\
Source densities are comparable at 6.75$\mu$m (1944 S$>$150$\mu$Jy sources
per square degree, Flores et al. 1998), 
15$\mu$m (2808 S$>$250$\mu$Jy sources per 
square degree, this paper) and  
5 GHz (1440 S$>$16$\mu$Jy sources per square degree,
Fomalont et al, 1991). Star-forming objects contribute respectively,
50\%, 73\% and 26\% of the extragalactic counts at 6.75$\mu$m,
15$\mu$m and 5 GHz. This suggests that the 60$\mu$m luminosity
density is strongly dominated by star-forming galaxies. 
The fraction of z $>$ 1 objects is
found to be $<$ 32\%, $<$ 43\% and $<$ 40\% of the extragalactic counts
 at 6.75$\mu$m, 15$\mu$m and 5 GHz, respectively. The 15$\mu$m survey is
found to be rather efficient in selecting high-redshift objects, since sources
with $I_{AB}\le$22.5 and $S_{15\mu}\ge$ 250 $\mu$Jy 
have a median redshift of 0.76 as compared to 0.59 for the  whole CFRS.\\

The 15$\mu$m survey, combined with radio and optical data, allows to
identify the most powerful star-forming objects (with $SFR$ larger than 
100 $M_{\odot} yr^{-1}$) in the field, at least
up to z = 1. Four such objects (0.7\%) are found 
among the 578 $I \le$ 22.5 galaxies,
and they contribute to 18\% of the SFR density. From their
UV or $[OII]_{3727}$ emission line properties, these objects cannot be
distinguished from galaxies with more modest rates of star formation. 
If we correct for extinction, assuming that their optical 
spectra from 2500 to 4000\AA\ is reddened by the standard galactic 
extinction curve (with $A_{V}$ derived from the $SFR_{IR}$ to $SFR_{UV}$
ratio), they would appear as
young starbursts with a moderate population of A stars ($W_{0}(H\delta)$=
3.5\AA ~in the combined spectrum). The 15$\mu$m sample contains highly 
reddened young starbursts as
well as a larger number of galaxies (S+SB) with lower SFRs, which all contain a
significant population of A stars (S+A galaxies). This is consistent with the 
scenario of strong starburst episodes followed by the last phases of the burst,
where the IR emission is still high due to dust heating by intermediate-mass
stars (M= 1-3 $M_{\odot}$, Lisenfeld et al, 1997).\\

Combination of 15$\mu$m and radio samples probably gives a good 
representation of the galaxy population in a very deep 60 $\mu$m survey. Interpolated  infrared (8-1000 $\mu$m) luminosities
based on fits of galaxy SEDs by local templates from radio to UV 
wavelengths, imply that 75\% (-40\%,$+$10\%) 
of the star formation rate density 
for $z \leq 1$ is hidden by dust. No evidence has been found for an  evolution
 of that fraction in the above redshift range. The global opacity of 
the Universe up to z = 1  ranges from $A_{V}$= 0.5 to 0.85.
A subsample of sixteen 15$\mu$m galaxies observed by the HST indicates 
that more than a third of the star formation hidden by dust is associated with 
interacting galaxies or mergers.\\
\\
Estimates of the SFR from UV fluxes carry some uncertainties because of 
two main reasons, one related to the UV light reprocessed by dust into FIR 
radiation, and the other, to the probably important contribution of AGN 
light at UV wavelengths. In spite of the small statistics of the sample 
considered here, 
our work allows a first glimpse of the true SFR density for $z \leq 1$.
From our careful data analysis, and the use of multi-wavelength data from 
radio to UV, we believe that the true SFR density lies within the
region delimited by the still large error bars in Figure 15. Our average
value, 75\% of the star formation rate density hidden by dust, is
consistent with:\\
- a one mag absorption at 2800\AA~ for the blue galaxy population
(see Tresse \& Maddox, 1998), and,\\ 
- less than 4\% of the galaxies are highly reddened, detected at 
15$\mu$m and at radio wavelengths and contribute to 50\% of the global IR
 luminosity density.\\
 
Although our values are lower by a factor 2.8 than those of Rowan Robinson
et al. (1997), they might be too high when the corresponding stellar mass
formed since z = 1 is compared to the
 present-day stellar mass density. This could raise an important question 
about the universality of the IMF, especially in the high SFR galaxies
detected by ISOCAM and  the VLA, and described in this paper. Studies  of
other  CFRS fields with the same multi-wavelength technique are needed 
to improve the source statistics, determine more accurately the SFR 
density, and study in more detail its redshift evolution.

{\it Acknowledgments:}

We thank Marc Sauvage and David Elbaz for useful discussions. We are 
also grateful to David
Schade who made available the HST images of the CFRS 1415+52 field 
in stamp format. Comments and criticisms from an anonymous referee has
led us to greatly improve the manuscript.

\acknowledgements

\clearpage

{\scriptsize
\begin{deluxetable}{rllllcccrlrr}
\tablenum{1}
\tablewidth{0pt}
\tablecaption{Optical counterparts of ISOCAM LW3 sources.}
\tablehead{
\colhead{ISO--} &\colhead{$\alpha_{2000}$} &\colhead{$\delta_{2000}$} &\colhead{CFRS} & \colhead{z$^1$} &
\colhead{I$_{AB}$} & \colhead{V$_{AB}$} & \colhead{K$_{AB}$} &
\colhead{d$^3$} & \colhead{P$^4$} & \colhead{Flux$^5$} &
\colhead{Error} 
}
\startdata
      \tableline \multicolumn{12}{|c|}{Catalogue 1: Objects with S/N $>$ 4 \& $P$ $<$ 0.02  } \nl \tableline
   0 &14:17:41.8 & 52:28:23.3 & 14.1157 &  1.150 &  20.54 &  22.45 &   --   &  0.65 & 0.000000$^7$ & 1653. &  57.\nl
   5 &14:17:41.9 & 52:30:23.2 & 14.1139 &  0.660 &  20.20 &  21.49 &  18.92 &  2.94 & 0.000383$^7$ &  562. &  48.\nl
   9 &14:17:40.4 & 52:28:21.1 & 14.1192 &   ---  &  23.49 &  24.40 &   --   &  0.92 & 0.012702 &  399. &  58.\nl
  13 &14:17:52.0 & 52:25:32.8 & 14.0855 &   ---  &  20.92 &  22.60 &   --   &  0.72 & 0.000998 &  487. &  69.\nl
  32 &14:17:56.6 & 52:31:58.6 & 14.0711 &   ---  &  21.44 &  22.29 &  20.26 &  1.50 & 0.006552 &  274. &  54.\nl
  42 &14:18:20.9 & 52:25:53.0 & 14.0098 &  star  &  14.66 &  16.44 &   --   &  1.53 & 0.000029 &  362. &  69.\nl
  43 &14:17:42.6 & 52:28:46.3 & 14.1129 &   ---  &  21.05 &  22.35 &  20.30 &  0.57 & 0.000694 &  209. &  52.\nl
  44 &14:17:34.9 & 52:27:51.0 & 14.1329 &  0.375 &  19.52 &  20.60 &   --   &  1.89 & 0.000294$^7$ &  347. &  52.\nl
  51 &14:17:24.3 & 52:30:24.0 & 14.1567 &  0.479 &  19.79 &  20.04 &  18.62 &  2.99 & 0.006937 &  459. &  57.\nl
  84 &14:17:45.8 & 52:30:31.2 & 14.1028 &  0.988 &  21.57 &  23.84 &  19.69 &  1.52 & 0.007463$^7$ &  295. &  50.\nl
 145 &14:18:07.1 & 52:28:37.3 & 14.0446 &   ---  &  20.00 &  21.33 &   --   &  0.31 & 0.000089 &  314. &  54.\nl
 206 &14:18:13.4 & 52:31:47.3 & 14.0272 &  0.668 &  20.51 &  21.70 &  19.12 &  1.51 & 0.003157 &  297. &  55.\nl
 233 &14:17:58.5 & 52:27:14.8 & 14.0663 &  0.743 &  20.88 &  22.34 &   --   &  2.51 & 0.002081 &  288. &  56.\nl
 282 &14:18:18.9 & 52:29:05.4 & 14.0138 &  star  &  15.77 &  16.80 &  15.43 &  5.66 & 0.000985 &  214. &  53.\nl
 294 &14:17:47.0 & 52:29:11.9 & 14.0998 &  0.430 &  20.58 &  21.84 &  19.06 &  2.44 & 0.008694$^7$ &  245. &  53.\nl
 303 &14:17:46.4 & 52:33:50.8 & 14.1006 &   ---  &  20.66 &  21.74 &   --   &  2.12 & 0.007003 &  270. &  53.\nl
 308 &14:17:26.7 & 52:32:20.9 & 14.1511 &   ---  &  20.71 &  21.61 &   --   &  0.78 & 0.000990 &  284. &  61.\nl
 369 &14:18:4.0 & 52:27:47.6 & 14.9154 &  0.812 &  21.57 &  23.06 &   --   &  3.25 & 0.000489$^7$ &  304. &  51.\nl
   \tableline \multicolumn{12}{|c|}{Catalogue 2: Objects with S/N $>$ 4 \& $P(d,I)$ $>$ 0.02 } \nl\tableline
   8 &14:17:23.8 & 52:27:49.3 & 14.1598 &   ---  &  19.84 &  21.00 &   --   &  8.87 & 0.061772 &  367. &  73.\nl
  24 &14:17:23.7 & 52:34:33.7 & 14.1582 &   ---  &  22.78 &  23.44 &   --   &  3.92 & 0.123123 &  341. &  77.\nl
 138 &14:17:26.3 & 52:32:51.5 & 14.1527 &   ---  &  23.00 &  24.24 &   --   &  5.54 & 0.268762 &  252. &  62.\nl
 139 &14:18:18.3 & 52:29:16.3 & 14.0150 &   ---  &  22.23 &  24.19 &  19.98 &  1.64 & 0.014691 &  284. &  57.\nl
 160 &14:17:37.4 & 52:31:41.9 & 14.1278 &   ---  &  22.08 &  28.36 &  20.31 & 11.04 & 0.448282 &  401. &  55.\nl
 190 &14:17:54.1 & 52:33:56.2 & 14.0779 &  0.578 &  22.01 &  23.10 &   --   &  2.94 & 0.039091 &  255. &  53.\nl
 195 &14:17:24.6 & 52:30:40.1 & 14.1569 &   ---  &  20.61 &  21.65 &  19.51 &  7.05 & 0.071948 &  364. &  61.\nl
 258 &14:17:41.1 & 52:30:22.3 & 14.1166 &  1.015 &  22.46 &  23.88 &  20.46 &  5.24 & 0.166123 &  188. &  47.\nl
 258 &           &            & 14.1178 &  9.999 &  22.47 &  24.66 &  20.24 &  6.38 & 0.237751 &  188. &  47.\nl
 278 &14:17:42.9 & 52:28:00.8 & 14.1103 &  0.209 &  22.33 &  22.64 &   --   &  7.84 & 0.306807 &  235. &  55.\nl
 278 &           &            & 14.1145 &   ---  &  22.64 &  25.72 &   --   &  7.72 & 0.365875 &  235. &  55.\nl
 278 &           &            & 14.1091 &   ---  &  22.02 &  22.79 &   --   & 10.70 & 0.412821 &  235. &  55.\nl
 278 &           &            & 14.1125 &   ---  &  22.62 &  25.75 &   --   &  8.45 & 0.415531 &  235. &  55.\nl
 296 &14:17:56.0 & 52:32:55.9 & 14.0741 &   ---  &  23.00 &  23.16 &   --   &  6.09 & 0.314943 &  207. &  49.\nl
 296 &           &            & 14.0743 &   ---  &  21.65 &  22.74 &   --   & 10.58 & 0.320898 &  207. &  49.\nl
 304 &14:17:27.3 & 52:32:09.6 & 14.1489 &   ---  &  20.91 &  20.54 &   --   &  9.14 & 0.147525 &  284. &  61.\nl
 307 &14:17:47.9 & 52:34:06.6 & 14.0975 &   ---  &  21.21 &  23.19 &   --   &  7.45 & 0.126166 &  252. &  55.\nl
 325 &14:18:12.5 & 52:32:48.4 & 14.0291 &   ---  &  22.19 &  24.51 &   --   &  3.88 & 0.077093 &  225. &  51.\nl
 361 &14:17:49.9 & 52:33:43.2 & 14.0909 &  0.978 &  22.34 &  24.30 &   --   &  2.12 & 0.026649 &  248. &  56.\nl
 372 &14:18:02.7 & 52:27:59.6 & 14.0557 &   ---  &  21.63 &  22.51 &   --   &  3.57 & 0.042434 &  225. &  52.\nl
 421 &14:17:52.5 & 52:35:13.4 & 14.9907 &   ---  &  22.95 &  19.41 &   --   &  7.78 & 0.447372 &  259. &  55.\nl
 434 &14:17:44.1 & 52:25:50.5 & 14.1070 &   ---  &  20.86 &  21.33 &   --   &  5.82 & 0.060281 &  277. &  65.\nl
 440 &14:17:39.3 & 52:28:46.1 & 14.1232 &   ---  &  22.06 &  23.19 &  20.05 &  3.91 & 0.070782 &  245. &  55.\nl
 440 &           &            & 14.1212 &   ---  &  20.20 &  20.65 &  20.30 &  8.87 & 0.081572 &  245. &  55.\nl
 457 &14:17:52.1 & 52:30:49.8 & 14.0846 &  0.989 &  21.81 &  23.15 &  20.34 &  2.97 & 0.034073 &  250. &  50.\nl
   \tableline \multicolumn{12}{|c|}{Catalogue 3: Objects with S/N $>$ 4 \&  without optical counterpart in I$_{AB}$ } \nl\tableline
 021 &14:17:23.2 & 52:32:58.3 & ---     &   ---  &  ---   &  ---   &  ---   &  ---  & 1.0000000 &  270. & 62.\nl
 202 &14:17:57.3 & 52:32:29.8 & ---     &   ---  &  ---   &  ---   &  ---   &  ---  & 1.0000000 &  209. & 49.\nl
 288 &14:17:53.6 & 52:33:44.6 & ---     &   ---  &  ---   &  ---   &  ---   &  ---  & 1.0000000 &  231. & 55.\nl
 439 &14:17:54.7 & 52:32:54.9 & ---     &   ---  &  ---   &  ---   &  ---   &  ---  & 1.0000000 &  248. & 53. \nl
    \tableline \multicolumn{12}{|c|}{Catalogue 4: Objects with 4 $>$S/N $>$ 3 \&$P$ $<$ 0.02} \nl\tableline
 171 &14:18:16.1 & 52:29:39.2 & 14.0198 &  1.603 &  20.04 &  20.21 &  19.86 &  1.52 & 0.002196 &  211. &  56.\nl
 183 &14:17:30.9 & 52:33:44.0 & 14.1400 &  star  &  15.61 &  16.13 &   --   &  5.59 & 0.000845 &  208. &  60.\nl
 228 &14:17:23.3 & 52:30:33.4 & 14.1609 &  star  &  18.93 &  20.83 &  18.30 &  2.54 & 0.002518 &  193. &  61.\nl
 243 &14:17:35.9 & 52:32:46.6 & 14.1302 &  0.548 &  20.85 &  21.72 &   --   &  3.19 & 0.018359 &  188. &  55.\nl
 310 &14:18:15.0 & 52:31:22.2 & 14.0227 &  0.772 &  20.84 &  22.08 &  19.47 &  3.42 & 0.020906 &  196. &  53.\nl
 326 &14:17:53.5 & 52:25:51.5 & 14.0818 &  0.899 &  21.02 &  22.14 &   --   &  2.74 & 0.010601 &  165. &  54.\nl
 351$^{6}$ &14:17:40.7 & 52:33:57.9 & 14.9504 &  ---   &  23.90 &  99.99 &   --   &  8.44 & 0.005986$^7$ &  228. &  59.\nl
 354 &14:17:47.8 & 52:32:52.8 & 14.0968 &  ---   &  23.39 &  24.88 &   --   &  1.07 & 0.015834 &  172. &  54.\nl
 396 &14:18:18.2 & 52:33:04.9 & 14.0151 &  ---   &  17.92 &  18.19 &   --   &  1.93 & 0.000647 &  185. &  53.\nl
 408 &14:17:40.5 & 52:27:14.7 & 14.1190 &  0.754 &  20.99 &  22.49 &   --   &  0.95 & 0.001837$^7$ &  185. &  56.\nl
 427 &14:17:59.7 & 52:26:01.4 & 14.0645 &   ---  &  22.44 &  24.40 &   --   &  9.72 & 0.000824$^7$ &  247. &  67.\nl
 449 &14:17:42.7 & 52:32:21.6 & 14.1117 &   ---  &  20.79 &  21.15 &   --   &  3.77 & 0.024364 &  213. &  54.\nl
 475 &14:17:53.9 & 52:31:37.0 & 14.0820 &  0.976 &  21.69 &  24.27 &  19.37 & 11.82 & 0.012867$^7$ &  203. &  53.\nl
    \tableline \multicolumn{12}{|c|}{Catalogue 5: Objects with 4 $>$S/N $>$ 3 \& $P$ $>$ 0.02} \nl\tableline
 137 &14:17:23.9 & 52:25:54.1 & 14.1597 &   ---  &  23.05 &  25.95 &   --   &  4.89 & 0.224187 &  284. &  93.\nl
 174 &14:17:45.3 & 52:29:47.6 & 14.1042$^1$ &  0.722 &  21.49 &  23.38 &  19.79 &  3.67 & 0.040133 &  197. &  50.\nl
 183 &14:17:30.8 & 52:33:54.2 & 14.1403 &   ---  &  21.48 &  24.21 &   --   &  4.96 & 0.071532 &  208. &  60.\nl
 228 &14:17:23.7 & 52:30:29.7 & 14.1591 &   ---  &  22.76 &  23.65 &  21.60 &  5.51 & 0.225442 &  193. &  61.\nl
 243 &14:17:36.0 & 52:32:55.4 & 14.1300 &   ---  &  23.25 &  24.15 &   --   &  5.65 & 0.328192 &  188. &  55.\nl
 276 &14:18:12.7 & 52:31:41.8 & 14.0287 &   ---  &  22.29 &  24.19 &  19.98 &  7.56 & 0.281069 &  175. &  51.\nl
 277 &14:17:24.7 & 52:26:56.2 & 14.1554 &   ---  &  21.90 &  23.07 &   --   &  5.51 & 0.120360 &  240. &  76.\nl
 341 &14:17:23.9 & 52:32:28.0 & 14.1576 &   ---  &  22.33 &  23.83 &   --   &  9.93 & 0.444487 &  214. &  64.\nl
 341 &           &            & 14.1596 &   ---  &  24.21 &  26.44 &   --   &  4.52 & 0.422721 &  214. &  64.\nl
 367 &14:17:25.7 & 52:31:32.2 & 14.1533 &   ---  &  22.42 &  23.19 &  21.81 &  7.46 & 0.299945 &  177. &  54.\nl
 377 &14:17:40.8 & 52:34:13.0 & 14.1170 &   ---  &  21.83 &  23.92 &   --   &  5.14 & 0.100134 &  233. &  60.\nl
 384 &14:17:42.6 & 52:29:54.3 & 14.1120 &   ---  &  22.60 &  23.57 &  20.78 &  6.25 & 0.251090 &  187. &  51.\nl
 396 &14:18:17.7 & 52:33:02.6 & 14.0157 &   ---  &  22.35 &  23.01 &   --   &  7.03 & 0.258736 &  185. &  53.\nl
 399 &14:18:05.8 & 52:32:10.2 & 14.0477 &   ---  &  23.01 &  24.73 &   --   &  4.36 & 0.177522 &  179. &  53.\nl
 408 &14:17:40.3 & 52:27:20.6 & 14.1195 &   ---  &  22.67 &  24.42 &   --   &  5.85 & 0.235036 &  185. &  56.\nl
 424 &14:18:12.3 & 52:29:45.4 & 14.0302 &   ---  &  20.87 &  22.20 &  19.22 &  5.07 & 0.046449 &  191. &  50.\nl
 426 &14:18:19.9 & 52:30:47.2 & 14.0112 &   ---  &  23.53 &  24.86 &  20.61 &  6.04 & 0.433874 &  160. &  53.\nl
 442 &14:17:38.5 & 52:34:43.2 & 14.1252 &   ---  &  21.91 &  21.79 &   --   & 10.02 & 0.347868 &  221. &  69.\nl
 449 &14:17:43.3 & 52:32:19.5 & 14.1100 &   ---  &  24.46 &  24.37 &   --   &  2.75 & 0.220019 &  213. &  54.\nl
 474 &14:17:31.7 & 52:29:11.0 & 14.1393 &   ---  &  22.23 &  23.67 &  21.28 &  2.42 & 0.031712 &  160. &  52.\nl
    \tableline \multicolumn{12}{|c|}{Catalogue 6: Objects with 4 $>$S/N $>$ 3 \& without optical counterpart in I$_{AB}$ } \nl\tableline
 231 &14:17:33.4 & 52:34:21.1 & ---     &   ---  &  ---   &  ---   &  ---   &  ---  & 1.0000000 &  209.& 59. \nl
 268 &14:17:31.4 & 52:25:37.5 & ---     &   ---  &  ---   &  ---   &  ---   &  ---  & 1.0000000 &  224.& 70. \nl
 348 &14:18:06.3 & 52:25:32.0 & ---     &   ---  &  ---   &  ---   &  ---   &  ---  & 1.0000000 &  193.& 67.    \nl
 359 &14:17:23.0 & 52:28:59.2 & ---     &   ---  &  ---   &  ---   &  ---   &  ---  & 1.0000000 &  234.& 73.   \nl
 483 &14:17:29.9 & 52:29:21.3 & ---     &   ---  &  ---   &  ---   &  ---   &  ---  & 1.0000000 &  167.& 53.    \nl
\enddata
\tablenotetext{1}{ The redshift for 14.1042 was given as z = 0.7217 by
  Lilly {\it et al.} (1995b). ``Redshifts'' for the
  four stars (denoted by `` star  '') were determined from additional
  spectra. ``---'' indicates that the redshift is unknown. }
\tablenotetext{2}{ ``----'' indicates that the magnitude is not available. }
\tablenotetext{3}{ Distance in arcsec between ISO source and optical or radio co
unterpart.}
\tablenotetext{4}{ Probability that the coincidence is by chance. }
\tablenotetext{5}{ Flux  within a 9\arcsec aperture in $\mu$Jy not corrected
for aperture effects (see text).}
\tablenotetext{6}{ Hammer {\it et al} (1995) and Fomalont {\it et al} (1992), 
comment that the fainter source (I$_{AB}\sim 23.9$) galaxy, which is 2 arcsec 
away, should be the optical counterpart. Unfortunately, no spectrum of the 
fainter galaxy is available.}
\tablenotetext{7}{ Radio-source with probability based on radio source counts.}
\end{deluxetable}
}

\clearpage

\begin{deluxetable}{cccc}
\tablenum{2}
\tablewidth{0pt}
\tablecaption{Counts of ISO catalogs at 15$\mu$m}
\tablehead{\colhead{} &\colhead{$S/N \geq 4$} & \colhead{$4 > S/N \geq 3 $} &\colhead{$S/N \geq 3$} }
\startdata
Total                &    41      &  37              & 78     \nl \tableline
\# $I_{AB} \leq 22.5$&   33       &  22              & 55     \nl
       with z        &   17       &  9               & 26     \nl
\# $I_{AB} > 22.5$   &    4       &  10              &  14    \nl
 No id               &    4       &  5               &  9     \nl \tableline
LW2 detection        &   10       &  7               & 17     \nl
radio detection      &    5       &  5               & 10$^1$ \nl \tableline
\enddata
\tablenotetext{1}{Radio sources with S$_{5GHz}\geq 16 \mu$Jy.}
\end{deluxetable}

\clearpage
\clearpage

{\tiny
\begin{table}[htbp]
\tablenum{3}
  \begin{center}
    \leavevmode
    \begin{tabular}{|l|rr|rr|rr|rr|rr|rr|rrc|l|} \hline
CFRS &\multicolumn{2}{|c|}{E} &\multicolumn{2}{|c|}{Sp} &\multicolumn{2}{|c|}{LI
NER} &\multicolumn{2}{|c|}{Seyfert2} &\multicolumn{2}{|c|}{SBL} &\multicolumn{2}
{|c|}{SBH} &  \multicolumn{3}{|c|}{Hybrid} & Class\\\hline
  Hyb         SB+S   Clas
        &$\chi^2$& Er &$\chi^2$& Er &$\chi^2$& Er &$\chi^2$&  Er &$\chi^2$& Er &
$\chi^2$& Er &$\chi^2$& Er &\%SBH+\%S&      \\ \hline
14.0227 & 12.69&  1.24&  3.18&  0.59&  2.47&  0.81&   0.97&  0.27&  2.21&  0.78&  0.37&  0.31&  0.19&  0.13&  9+1&   SB+S \\
14.0272 & 12.11&  1.69&  4.45&  0.93&  2.54&  0.88&   1.16&  0.45&  1.58&  0.60&  0.41&  0.32&  0.29&  0.20&  8+2&   SB+S \\
14.0276 &  4.22&  0.31&  0.75&  0.22&  0.75&  0.18&   0.51&  0.17&  4.10&  0.85&  1.32&  0.20&  0.15&  0.12&  8+2&   SB+S \\
14.0573 &      &      &      &      &      &      &       &      &      &      &      &      &      &      &     &   Seyfert 1 \\
14.0663 &  8.11&  0.61&  1.51&  0.70&  1.58&  0.26&   0.59&  0.04&  2.18&  0.84&  0.47&  0.06&  0.51&  0.10&  9+1&   SBH \\
14.0727 & 11.90&  0.95&  4.07&  1.25&  2.89&  0.38&   1.41&  0.30&  0.70&  0.30&  0.55&  0.16&  0.23&  0.19&  8+2&   SB+S \\
14.0779 & 16.16&  2.25& 21.11& 18.57&  6.07&  1.31&   0.55&  0.17&  1.73&  0.54&  0.63&  0.30&  1.40&  0.67&  9+1&   Seyfert 2 /SBH \\
14.0818 & 13.20&  0.89&  3.0&6  0.15&  4.10&  0.54&   1.17&  0.22&  2.27&  2.17&  0.48&  0.32&  0.23&  0.11&  8+2&   SB+S  \\
14.0820 &  2.94&  2.21&  0.62&  0.19&  1.69&  1.85&   2.13&  0.22& 16.92&  4.31&  6.59&  0.99&  0.64&  0.26&  1+9&   S\\
14.0846 &  8.94&  0.74&  2.17&  0.22&  3.65&  1.16&   0.45&  0.19&  4.85&  3.24&  0.83&  0.29&  0.35&  0.22&  4+6&   SB+S /Seyfert2 \\
14.0854 &  3.46&  0.43&  0.73&  0.22&  0.94&  0.40&   0.21&  0.10&  6.47&  1.27&  1.42&  0.11&  0.06&  0.13&  8+2&   SB+S /Seyfert2 \\
14.0909 &  5.13&  0.97&  1.51&  0.33&  1.55&  0.24&   0.15&  0.01&  6.36&  1.05&  1.67&  0.21&  1.46&  0.56&  4+6&   Seyfert 2 \\
14.0937 & 15.61&  9.20&  3.63&  0.72& 10.81&  4.20&   1.53&  0.16& 12.68&  6.98&  8.53&  3.37&  2.79&  1.11&  1+9&   Seyfert 2 \\
14.0998 &  7.89&  1.75&  4.42&  4.22&  1.42&  0.19&   0.86&  0.18&  2.92&  0.23&  0.51&  0.22&  0.31&  0.20&  6+4&   SB+S \\
14.1028 &  3.13&  0.43&  0.86&  0.31&  0.66&  0.16&   0.25&  0.12&  6.23&  0.41&  1.47&  0.15&  0.54&  0.24&  4+6&   Seyfert 2 /SB+S \\
14.1041 & 23.07&  6.23& 33.67& 72.26&  6.21&  1.72&   1.45&  0.11&  1.59&  0.93&  7.34&  3.11&  1.81&  1.08&  8+2&   Seyfert 2 /SB+S \\
14.1042 &  8.29&  0.75&  3.01&  1.18&  1.43&  0.16&   0.45&  0.13&  4.01&  0.25&  0.95&  0.18&  0.38&  0.15&  8+2&   SB+S /Seyfert2 \\
14.1103 &      &      &      &      &      &      &       &      &      &      &      &      &      &      &     &     HII \\
14.1129 & 13.09&  2.76&  2.77&  0.63&  3.53&  1.99&   0.97&  0.47&  1.23&  0.70&  0.54&  0.28&  0.49&  0.31&  7+3&   SB+S  \\
14.1139 &  8.83&  0.24&  2.05&  0.12&  1.69&  0.09&   0.66&  0.08&  1.36&  0.05&  0.28&  0.06&  0.81&  0.07&  9+1&   SBH  \\
14.1190 &  8.57&  0.79&  1.34&  0.27&  1.59&  0.58&   0.80&  0.21&  2.01&  0.94&  0.37&  0.17&  0.45&  0.13&  9+1&   SBH \\
14.1302 &      &      &      &      &      &      &       &      &      &      &      &      &      &      &     &   Seyfert 1 \\
14.1303 &      &      &      &      &      &      &       &      &      &      &      &      &      &      &     &   QSO \\
14.1329 & 18.24&  1.11&  8.25&  0.13&  3.01&  0.11&   3.29&  0.10&  3.47&  0.14&  2.91&  0.42&  4.62&  0.63&  9+1&   SBH /Seyfert2 \\
14.1567 &      &      &      &      &      &      &       &      &      &      &      &      &      &      &     &   Seyfert 1 \\
14.9025 &      &      &      &      &      &      &       &      &      &      &      &      &      &      &     &   Liner \\
14.9154 & 25.58&  0.27&  6.54&  0.18& 12.43&  0.23&   1.52&  0.08& 12.15&  0.23&  5.43&  0.12&  5.47&  0.19&  4+6&   Seyfert 2 \\ \hline
    \end{tabular}
    \caption{$\chi^{2}$ values for all galaxies of the 15$\mu$m and radio
samples and classifications.}
  \end{center}
\end{table}
}
\clearpage


\clearpage
\begin{table*}
\tablenum{5}
\caption{Global Luminosities [$10^10*L_\odot$] of the CFRS 1415+52 field}
\label{table5sfr}
\leavevmode
  \begin{center}
    \begin{tabular}{|c|cc|cc|cc|} \hline
   & \multicolumn{2}{|c|}{15$\mu$m-Radio}& \multicolumn{2}{|c|}{no-detected 15$\mu$m -Radio}
&\multicolumn{2}{|c|}{CFRS1415+52 } \\ \hline
                & L$_{2800}$   & L$_{IR}$  & L$_{2800}$ & L$_{IR}$     & L$_{2800}$ & $L_{IR}$\\ \hline
Upper limit     &  29$\pm$6 & 1778$\pm$607 & 217$\pm$29 & 2161$\pm$214 &246$\pm$35 &{\bf 3939$\pm$821 }\\ \hline
AGN contrib.    &   4$\pm$1 &  680$\pm$291 &  17$\pm$ 2 &   84$\pm$ 27 & 21$\pm$3  &      724$\pm$318  \\ \hline
"old stars" contrib&   1$\pm$0 &  115$\pm$ 28 &  25$\pm$ 3 &   44$\pm$ 16 & 26$\pm$3  &      159$\pm$ 44  \\ \hline
Lower limit     &  24$\pm$5 &  984$\pm$288 & 175$\pm$24 &  511$\pm$ 71 &199$\pm$29 &{\bf 1495$\pm$359 }\\ \hline
  \end{tabular}
  \end{center}
\end{table*}

\clearpage

\end{document}